\documentclass[%
 aps,
 amsmath,amssymb,
 reprint,
]{revtex4-1}

\usepackage{graphicx}
\usepackage{dcolumn}
\usepackage{bm}
\usepackage{hyperref} 
\usepackage{verbatim} 
\usepackage[utf8]{inputenc}
\usepackage[T1]{fontenc}
\usepackage{mathptmx}
\usepackage{multirow}
\usepackage{float}
\usepackage[english]{babel}
\usepackage[utf8]{inputenc}

\usepackage{xcolor}

\preprint{APS/123-QED}
\begin{document}

\title{Lattice dynamics in the conformational environment of multilayered hexagonal boron nitride (h-BN) conveys to peculiar infrared optical responses} 

\author{Luigi Cigarini$^{1}$} 
\author{Michal Novotn\'{y}}%
\author{Franti\v{s}ek Karlick\'{y}$^{2}$}%

\affiliation{Department of Physics, Faculty of Science, University of Ostrava, 30. dubna 22, 701 03 Ostrava, Czech Republic.}
\email{$^{1}$luigi.cigarini@osu.cz\\$^{2}$frantisek.karlicky@osu.cz}

\date{\today}

\begin{abstract}
Stacking mismatches in hexagonal boron nitride (h-BN) nanostructures affect their photonic, mechanical, and thermal properties. To access information about the stacked configuration of layered ensembles, highly sophisticated techniques like X-ray photoemission spectroscopy or electron microscopy are necessary. 
Here, instead, by taking advantage of the geometrical and chemical nature of h-BN, we show how simple structural models, based on shortened interplanar distances, can produce effective charge densities. 
Accounting these in the non-analytical part of the lattice dynamical description makes it possible to access information about the composition of differently stacked variants in experimental samples characterized by infrared spectroscopy. 
The results are obtained by density functional theory and confirmed by various functionals and pseudopotential approximations. 
Even though the method is shown on h-BN, the conclusions are more general and show how effective dielectric models can be considered as valuable theoretical pathways for the vibrational structure of any layered material.
\end{abstract}

\maketitle

\section{Introduction}
\label{sec:Introduction}
Layered materials are widely studied due to their ability to form stable 2D crystals. Exotic and unexpected properties like astounding electronic conductivity or unusual optical response \cite{roldan2017theory} can arise in some materials when reduced to their 2D forms\cite{kumari2016tuning}. 
Hexagonal boron nitride (h-BN) is considered a counterpart to graphene as its structure is almost identical, yet h-BN is a semiconductor with a relatively large electronic (6.08 eV) \cite{cassabois2016hexagonal, kolos2019accurate, henck2017direct} and optical (5.69 eV) \cite{kolos2019accurate,henck2017direct,Doan2016} band gap in bulk. Single h-BN layers can be efficiently used as dielectric layers in graphene nanostructures, electronics, or for lubrication, or directly replace other materials for high-temperature applications \cite{Yankowitz2019}. The crystalline structure of h-BN has been addressed in previous theoretical studies \cite{serrano2007,catellani1987bulk,park1987band,furthmuller1994ab,xu1991calculation,ohba2001first,liu2003structural,constantinescu2013stacking, bjorkman2012van, kolos2019accurate}, yet the question of stability for different possible stacking orders has not been answered to a satisfactory level \cite{liu2003structural,constantinescu2013stacking,cusco2018, marom2010stacking, kolos2019accurate}.

\begin{figure}[ht]
\centering
  \includegraphics[height=2.7cm]{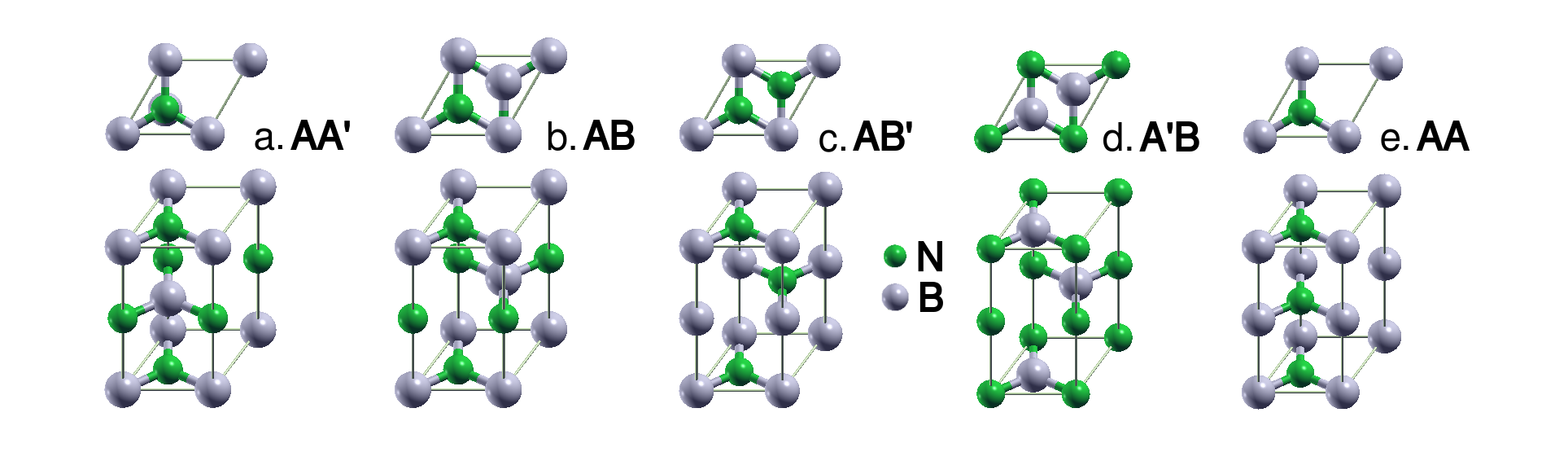} 
  \caption{Five different stacking orders in hexagonal boron nitride (top and axonometric views).}
  \label{fig:stackings}
\end{figure}

From a geometrical point of view, it is clear that a wide variety of stacking arrangements are possible for the of contiguous planes and several of them have been reported experimentally \cite{Warner2010,JCPDS_45_0895,JCPDS_73_2095,gilbert2019alternative,ni2019soliton,park2020one}. Five possible arrangements are shown in Fig.\ref{fig:stackings}. The AA' and AB types of stacking orders have been observed by atomic resolution imaging \cite{Warner2010,park2020one} and are known to be stable.  
H-BN is generally considered to be in the AA' stacking. New experimental routes \cite{gilbert2019alternative}, recently showed how to produce a purely AB stacked material. AA and A'B have been reported by theoretical calculations to be unstable \cite{liu2003structural}, nevertheless they have been found in traces in experimental measurements \cite{JCPDS_45_0895,JCPDS_73_2095}, while at the same time the actual amount of AB' stacking in h-BN samples remains unclear \cite{qi2007planar,liu2003structural,marom2010stacking,ni2019soliton,henck2017stacking}. The uncertainty is increased by the wide variability in experimental data obtained from infrared optical response \cite{Geick1966,hidalgo2013high,ccamurlu2016modification,mukheem2019boron,chen2017thermal,wang2003synthesis,andujar1998plasma}, Raman optical activity and  \cite{Geick1966,Nemanich1981,Reich2005,arenal2006raman}, photoluminescence spectroscopy \cite{museur2008defect, taylor1994observation, museur2008near}. 
 This may be due to a variety of different sources (among the others: possible poly-crystallinity, presence of amorphous regions, impurities), including stacking defects and variability in the amount of differently stacked regions in the probed specimens. Although many stacking combinations are possible, only two \cite{Geick1966} or three \cite{mosuang2002relative} of them have been considered in different theoretical studies related to the interpretation of experimental results. Until now, there has been no reported simple way to unravel the stacking composition of samples.

An accurate description of the lattice dynamics and stacking composition of bulk h-BN would produce a better explanation of experiments \cite{jung2015vibrational} and might contribute to achieving higher control over the crystalline aspects of the produced samples, as well as finer tuning of the final properties of interest, a precise value of thermal conductivity \cite{jiang2018anisotropic, lindsay2012theory,zhou2014high,jo2013thermal} or optical and photonic response \cite{lazic2019dynamically, vuong2016phonon, wang2015layer, kim2013stacking, henck2017stacking}, control over the materials in nanostructuring \cite{slotman2014phonons,wirtz2003ab,arenal2006raman,xu2019optomechanical,jung2015vibrational}. An important example of how stacking behaviour may have a practical impact is in the recent discovery that the local strain of the h-BN lattice framework is related to the single photon quantum emission properties \cite{hayee2020revealing}. 
But, specifically, a great anisotropy between the parallel and perpendicular plane dielectric constants has been reported for bulk h-BN in theoretical \cite{xu1991calculation} and experimental studies \cite{Geick1966}, 
yet theoretical models failed to reproduce the measured results \cite{ohba2001first}.
As we thoroughly describe in Section \ref{subsec:TheorDescr}, the different nature of the physical forces involved (covalent in-plane interactions and the inter-plane van der Waals forces) makes striving for a uniform theoretical interpretation extremely difficult.

Theoretical descriptions of the lattice dynamics of solids usually utilize density functional theory (DFT) \cite{kohn1965self}. 
Particularly, local density approximation (LDA) exchange-correlation (xc) functionals are often reported to be the best compromise for a consistent description of h-BN \cite{hamdi2010ab, ahmed2007first,mosuang2002relative,cusco2018}, nevertheless, satisfactory theoretical interpretation has not yet been reported on the paradoxical success of LDA.
Janotti \textit{et al.} \cite{janotti2001first} performed an all electron DFT calculations on bulk h-BN. The comparison of LDA results with the more sophisticated generalized gradient approximation (GGA) did not produce substantial improvement in the interpretation of the stacking interaction. The same conclusion was reported by Mosuang \textit{et al.} \cite{mosuang2002relative} with the use of norm-conserving (NC) pseudopotential (PP) approximations.
LDA theoretical implementations have produced results in a relatively satisfactory agreement with experimental values for the calculations of lattice constants, bulk moduli and cohesive energies of this material by employing any norm-conserving (NC) \cite{xu1991calculation,mosuang2002relative}, ultrasoft (US) \cite{furthmuller1994ab} or projector augmented-wave (PAW) \cite{ooi2005electronic} pseudopotential (PP) approximations.
Kim \textit{et al.} \cite{kim2003first} reported a significant improvement in the calculation of lattice constants using Troullier–Martins \cite{troullier1991efficient} NC PP, generated within the GGA correction, but implemented in an LDA xc functional theory. 
As a consequence, the lattice dynamics of bulk h-BN and the issue of relative stability of the different stacking orders until the present work, has been examined mainly by LDA approaches, employing US PP \cite{kern1999ab, yu2003ab, liu2003structural, ohba2001first} and NC PP \cite{cusco2018, serrano2007, karch1997ab}. GGA approaches \cite{topsakal2009first} never showed noticeable improvement before. 
A new promising candidate to resolve these issues, the strongly constrained and appropriately normed (SCAN) \cite{sun2016accurate,VASP_SCAN} functional has been advanced to deal efficiently with diversely-bonded materials (including covalent and van der Waals interactions). 

In this work, we present a comprehensive study of the lattice dynamics of bulk h-BN comparing five different possible stacking variants (within different levels of DFT functionals, PPs and vdW corrections). 
We employ an original methodology, based on a separate description of the analytical (inter-atomic force constants) and non-analytical parts (Born effective charges and dielectric tensors) of the dynamical matrix and we propose a simple, but effective, approach for the non-analytical calculations. 
Our chosen methodology of utilizing different levels of theory (LDA, GGA, SCAN, US-PP, NC-PP, PAW-PP) is not merely empirical, but arises from a careful consideration of the physics responsible for nuclear motion and dielectric dynamics. Using the available experimental data (without insights into their conformational composition), we are able to asses the quality of our calculations (Section \ref{subsec:AccuracyModel}) and derive a semi-empirical method (Section \ref{subsec:EmpDeriv}) which is potentially able to determine the conformational composition of the given samples. Overcoming the possible experimental difficulties, future developments (comprehensive data sets and powerful theoretical implementations) could permit a numerical shift in the present model and result, then, in an accurate interpretation of the measurements.
\\
\\
\\
\section{Methods}
\label{sec:Methods}

\subsection{Lattice dynamics}
\label{subsec:LatDyn}

We first analyzed the structure of the considered systems by means of well established phonon dynamical theories, within the theoretical scheme of the harmonic approximation. The vibrational movements, in periodic systems, are defined by a wave vector $\textbf{q}$ and a mode number $\nu$. For each wave vector and mode it is possible to determine an energy (vibrational frequency $\omega_{\nu}(\textbf{q})$) and displacements vectors $\textbf{u}^{\alpha}_{s, \nu}(\textbf{q}) $ by solving the phononic eigenvalue problem \cite{giannozzi2005density,wilson1941some,wilson1955molecular} deriving the interatomic force constants (IFC) by means of a finite displacements (FD) method.

This analytical approach is not sufficient to account for long-range Coulombic forces which originate in real crystal structures and give rise to a longitudinal - transverse optical (LO-TO) modes splitting. It is necessary to introduce a non-analytical (NA) correction term in the computed dynamical matrices. This approach was first derived by Cochran and Cowley \cite{cochran1962dielectric} based on the Born and Huang \cite{born1954dynamical} theoretical framework, successively adapted by Pick \textit{et al.} in the form which is currently implemented in the modern theories \cite{pick1970microscopic,ohba2001first}. 
It is based on a separation of the dynamical matrix into two different terms. The eigenvalue problem can then be reformulated, in real space, for the $\alpha$ direction (without noting the explicit dependence over $\textbf{q}$ of $\omega$, $W$ and $N$):

\begin{equation}
     \omega_{\alpha}^{2} W= \bigg(   N + \frac{4 \pi}{\epsilon^{(\infty)}_{\alpha \alpha}} Z   \delta_{\alpha \beta}   Z^{T}   \bigg) W   ,
\end{equation}

where the W matrix contains the displacement vectors {$\textbf{u}_{s}$} of the atoms {of mass $\mu_{s}$ (labeled by index $s$)}:
$W_{s}^{\alpha}  =  \sqrt{\mu_{s}}  \cdot  \textbf{u}_{s}^{\alpha}$. 
The $N$ matrix contains information about the interatomic force constants (IFC) of the system: 
$ N_{st}^{\alpha \beta} (\textbf{q})   \propto  {{\tilde{C}}_{st}^{\alpha \beta}}(\textbf{q})/({\sqrt{\mu_{s}  \mu_{t}}} )$
where $t$ is a second index for nuclei. 
$\tilde{C}_{st}^{\alpha \beta} (\textbf{q}) = \sum_{l} e^{-i\textbf{q}R_{l}} {C}_{st}^{\alpha \beta} (R_{l})$ is the analytical part of the dynamical matrix of the system.
${C}_{st}^{\alpha \beta} (R_{l})$ are the elements of the IFC, defined as:

\begin{equation}
{C}_{st}^{\alpha \beta} (l,m)  =  \frac{\partial^2 E}{\partial \textbf{u}_{s}^{\alpha} (l) \partial \textbf{u}_{t}^{\beta}(m)}  ,
\label{eq:IFC}
\end{equation}

where $E$ is the total energy of the system, $R_l$ is the nuclear coordinates matrix in the $l$ unit cell, $l$ and $m$ are indices of different unit cells. The explicit dependency of ${C}_{st}^{\alpha \beta}$ on couples of unit cell $(l,m)$ is intended, in Eq. \ref{eq:IFC}, to account for numerical implementations of IFC calculation. In the NA part, the $Z$ matrix ($ Z^{\alpha  \beta}_{s}  = {Q^{\alpha  \beta}_{s}}/{ \sqrt{\mu_{s}}}
\label{z}$) contains elements of the $Q$ matrix of the effective charges (later defined in Equation \ref{eq:pol}).

Giannozzi \textit{et al.} \cite{giannozzi1991ab}, Ohba \textit{et al.} \cite{ohba2001first,ohba2008erratum} and Gonze \textit{et al.} \cite{gonze1997dynamical} obtained effective charges and dielectric tensors by density functional perturbation theory (DFPT) and directly related them to the LO-TO splitting of the phononic modes and absorption activities. In a slightly different approach, here we consider the charges $Q^{* \alpha   \beta}_{s}$ \cite{note_star} and dielectric tensors $\epsilon$ calculated by DFPT as numerical counterparts of the $Q$ matrix elements and $\epsilon^{(\infty)}$ matrices of fictitious systems (the problem is shifted to the ideal creation of them and understanding the underlying relation). By constraining the cell in a single direction (perpendicular to the BN planes), we propose the aforementioned systems and thereby approximate the dynamical behavior of the dielectric dispersion matrix.


\subsection{Infrared optical response}
\label{subsec:VibSpec_Methods}
Porezag \textit{et al.} \cite{porezag1996infrared} applied the pioneering work of Wilson \textit{et al.} \cite{wilson1955molecular} to derive an expression for the absorption activity of phononic modes. The infrared (IR) absorption activity of a mode $\nu$ is:

\begin{equation}
    I^{IR}_{\nu}   =  \frac{\textit{n}\pi}{3c}   \bigg|\frac{\partial \boldsymbol{\mu}}{\partial\textbf{q}} \bigg|_{\nu}^{2}   ,
    \label{eq:IR_int}
\end{equation}

where $\textit{n}$ is the particle density, $c$
is the velocity of light and $\boldsymbol{\mu}$
is the electric dipole moment of the system. The expression is reformulated by defining the Born effective charges $Q^{*  \alpha   \beta}_{s}$ \cite{giannozzi2005density}:

\begin{equation}
    I^{IR}_{\nu}    =     \sum_{\alpha}  \bigg| {\sum_{s,\beta}}  Q^{*  \alpha   \beta}_{s}    \textbf{u}_{s, \nu}^{\beta}   \bigg|^{2}   .
    \label{eq:IR_int_born}
\end{equation}

Our use of DFPT in the calculation of the the NA dynamical parameters (e.g. $Q^{*  \alpha   \beta}_{s}$, which takes also part, Eq. \ref{eq:IR_int_born}, in the IR activity) differs conceptually from how the theory was originally conceived. Here we propose it as a numerical tool, which permits to relate uniquely every (possibly fictitious) charge density function to an absorption spectrum.
We point out, that the $\Gamma$ point eigenvectors of the IR active modes can be directly compared with the experimental peak frequencies, and we notice that a natural broadening of the spectral lines, in a Lorentz function shape, occurs due to finite lifetime of the phonon collective excitations as a consequence of scattering processes.

\subsection{Semi-empirical description of the dielectric dynamics}
\label{subsec:TheorDescr}
\begin{figure}[!ht]
\centering
  \includegraphics[width=8.5cm]{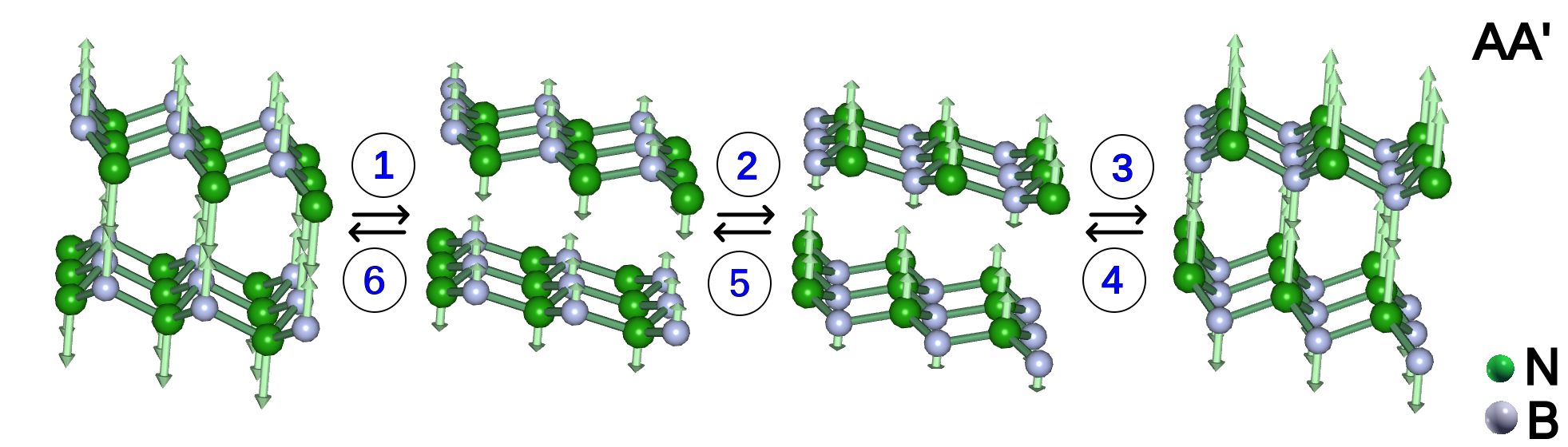}
  \caption{
  Four different dynamical "snapshots" of the IR active  $A_{2u}$ vibrational mode of the AA' configuration of h-BN (periodic movement in time is indicated by numbers and black arrows), which correspond to an ammonia-like inversion movement. Green arrows indicate the direction and magnitude of the displacement eigenvectors of the atom from its equilibrium position.}
  \label{fig:a2u}
\end{figure}

In this paragraph, we present a short qualitative description that introduces our original approach. 
We begin with a molecular orbital (MO) depiction of the multilayered h-BN system, where the $sp^3$ orbitals related to the nitrogen atoms bear an electronic "lone pair". In this picture, the configuration of the orbitals and the relevant electronic clouds resonate between two limit forms in the upper and lower sides of the h-BN plane (ammonia-like inversion movement, see Figure \ref{fig:a2u}). 
Strong instantaneous dipoles are responsible for the van der Waals stacking interaction, in a continuous fluctuation of the charge density (and dielectric dispersion) between the two sides of the h-BN plane.

 Cochran and Cowley \cite{cochran1962dielectric} elegantly derived the following relation between the fluctuations of the dielectric dispersion involved in a vibrational mode related to the lattice dynamics of it (the ratio on the left side of Equation \ref{eq:omega_epsilon} between the frequencies $\omega_{\nu}$ of phonons and dielectric dispersion frequencies $\Omega_{\nu} $) and the quotient of two dielectric constants of the system (on the right side of Equation \ref{eq:omega_epsilon}):
\begin{equation}
    \prod_{\nu=4}^{3n}     \bigg(     \frac{\omega^{\alpha}_{\nu}}{\Omega_{\nu}}      \bigg)^{2}         =        \frac{\epsilon_{\alpha    \alpha}^{(0)}}{\epsilon_{\alpha    \alpha}^{(\infty)}}  ,
    \label{eq:omega_epsilon}
\end{equation}
where $\nu$ is the index denoting different normal vibrational modes, $n$ is the total number of atoms in the asymmetric unit, $\alpha$ is a space direction ($x \lor y \lor z$), $\epsilon^{(\infty)}$ and  $\epsilon^{(0)}$ are respectively  the high frequency and static  dielectric tensors. Recall that the high frequency dielectric tensor $\epsilon^{(\infty)}$ depends only on the electronic susceptibility, i.e. just on charge density fluctuations, while the static  dielectric tensor $\epsilon^{(0)}$ depends also on the lattice dynamics \cite{cochran1962dielectric, born1954dynamical}. 

Ideally, two distinct values of frequency, $\omega_{\nu}$ and $\Omega_{\nu} $, could be associated to a normal mode (Equation \ref{eq:omega_epsilon}), but in a mode with zero component of polarization in the $\alpha$ direction $\Omega_{\nu} = \omega^{\alpha}_{\nu}$.
However, the dielectric dispersion frequencies 
(rigorously defined also in \cite{cochran1962dielectric}) could be thought of as the frequencies of vibration of a fictitious system in which the dielectric dynamics is reproduced exactly. The dielectric dynamics can be described as dependent on an ideal "apparent charge density" $\rho_{(Q)}(\textbf{r})$, $\textbf{r}  \equiv (x,y,z)$, acting in determining the effective polarization of the crystal. 

Accounting this idealization, the total polarization is composed, in addition to the simple electronic polarization, by an ionic term $Q^{T} U$ \cite{born1954dynamical, cochran1962dielectric}:
\begin{equation}
    \textbf{P}^{\alpha}   = \sum_{s}   Q^{\alpha \beta}_{s} \textbf{u}^{\beta}_{s}  +  \chi^{\alpha \beta}   \textbf{E}^{\alpha} ,
    \label{eq:pol}
\end{equation}
where $U$ is the matrix of the ionic displacements $\textbf{u}^{\alpha}_{s}$,  
$\chi$ is the electronic susceptibility matrix, $\textbf{E}^{\alpha}$ are the real space components of the macroscopic electric field vector. 
The $Q$ matrix (dimension $3n \times 3$, where $n$ is the total number of atoms in the unit cell) is an expression of the "apparent charge density" $\rho_{(Q)}(\textbf{r})$. Understanding this relation ($\rho_{(Q)}(\textbf{r}) \rightarrow Q$) is one of the purposes of this study (for which we propose DFPT as a numerical tool, applied to fictitious, shortened systems).

We note that a given geometrical configuration and ground state charge density produce a unique "apparent charge density" $\rho_{(Q)}(\textbf{r})$. This means that not only two systems with different geometrical conformations, but even two systems with the same geometrical conformation and different ground state charge density distributions (e.g. calculated with different DFT xc functionals, or different pseudopotential approximations) can be thought as simulating diverse systems or as simulating varying average phases in a charge density fluctuation process, like that involved in a phononic mode vibration (like the $A_{2u}$, qualitatively depicted in Figure \ref{fig:a2u}).


By shortening the $c$ lattice parameter (orthogonal to the h-BN planes), we can produce a systems in which the frequency of vibration of the dielectric dispersion $\Omega_{\nu}^{fict.}$ associated to the $A_{2u}$ mode, approaches progressively the $\Omega_{\nu}$ of the real system \cite{cochran1962dielectric}. 
By determining an optimal shortening percentage, the uniquely produced $Q^{* \alpha   \beta}_{s}$ and $\epsilon$ effectively model the $Q$ matrix elements and $\epsilon^{(\infty)}$ matrices of real systems (NA parameters of the dynamical matrices), respectively, i.e. reproduce the actual ratio between the intensities of the two IR active absorptions. 
This optimization 
can also be observed when looking at the charge densities, 
as seen in Appendix: the "apparent charge density", in the hypothetical polar cones of nitrogen, of the optimally shortened fictitious system mimics the effective function $\rho_{(Q)}(\textbf{r})$ of the real system.

\subsection{Computational details}
\label{subsec:CompDetails}
The calculations are performed by means of the Quantum Espresso (QE) \cite{giannozzi2009quantum} package and VASP code \cite{ Kresse_PhysRevB_54_1996, Kresse_Comp_Mat_Sci_6_1996, Kresse_PhysRevB_49_1994, Kresse_PhysRevB_47_1993}, using a periodic density functional theory (DFT) \cite{kohn1965self} in a plane waves (PW) basis set \cite{giannozzi2009quantum, Kresse_PhysRevB_59_1999, Kresse_PhysRevB_54_1996,Kresse_PhysRevB_49_1994,Kresse_PhysRevB_47_1993,Kresse_Comp_Mat_Sci_6_1996}. We treat the exchange-correlation (xc) potential within the electronic Hamiltonian in different manners: Local Density Approximation (LDA) \cite{hohenberg1964inhomogeneous}, Generalized Gradient Approximation (GGA) \cite{perdew1996generalized_1} in the well-known Perdew–Burke–Ernzerhof (PBE) \cite{perdew1996generalized} formulation and the recently proposed Strongly Constrained and Appropriately Normed (SCAN) \cite{sun2016accurate,VASP_SCAN} functional. We employ the Vanderbilt Ultrasoft (US) PP \cite{vanderbilt1990soft}, Troullier–Martins FHI Norm-conserving (NC) PP \cite{fuchs1999ab, hamann1979norm,hamann2013optimized,troullier1991efficient,hamann2017erratum} methods and the Projector Augmented-Wave (PAW) method of Blöchl \cite{Blochl_PhysRevB_50_1994}. US and PAW PP, in spite of a lower cutoff value in the PW basis set and a sensible reduction of computational time, do not guarantee the conservation of the norm for the resulting wave functions in comparison with all electron calculations, especially  outside the core-shell regions of atoms. This drawback could particularly affect properties like phonons, which involve the calculation of interactions at longer than optimal distances with respect to covalent bonds. For this reason, in some of our implementations, we recalculate the charge densities after each diagonalization step, using dramatically higher (8-20 times) values of cutoff for the PW (see values of $E_{cut.}\rho$ with respect to $E_{cut.}$ in Electronic Supplementary Information (Table \ref{tab:calc_param}). 

To account for the lack of DFT to properly calculate weak interactions such as the interplanar van der Walls forces, we supplemented our DFT with empirical dispersion correction methods: Grimme-D2 \cite{Grimme2006} and D3 with Becke-Johnson damping (D3BJ) \cite{Grimme2011} methods, Tkatchenko-Scheffler method (TS) \cite{Tkatchenko2009} as well as the SCAN+rVV10 \cite{VASP_SCANrVV10} functional.

For each method we performed structural relaxations and phonon dispersion calculations. The calculation parameters and methods are summed up in Electronic Supplementary Information (Table \ref{tab:calc_param}). We employ different fitting techniques to extrapolate the optimal structural parameters and the resulting values used in subsequent calculations are reported too, in Electronic Supplementary Information (Table \ref{cell_par}).
The IFC have been calculated, after geometrical optimization, by means of a FD approach. From here on, the symmetries of the structures have been imposed as reported in E.S.I., Tab. \ref{tab:coordinates}. For the FD self-consistent field (SCF) calculations we employed VASP in conjunction with the Phonopy code \cite{phonopy} and the PW and FD algorithms included in QE. 
The DFPT Born effective charges and DFPT dielectric tensors \cite{ohba2001first,ohba2008erratum,giannozzi1991ab} (here numerical expressions of $Q$ matrix elements and $\epsilon^{(\infty)}$ matrices) are obtained setting a threshold for self-consistency of $1.36\cdot10^{-12}$ eV. 
The reported IR spectra are calculated by applying a Lorentz broadening function to the absorption intensities of the vibrational modes in order to have an half width at half maximum of 10 cm$^{-1}$. We separately applied the broadening functions to the single phonon spectral lines.

The potential energy surfaces (PESs) presented in Figures \ref{fig:PESstable} and in the Electronic Supplementary Information (Figure \ref{ABprime_PES}) have been obtained by means of the QE package with single point SCF calculations. These have been performed on Monkhorst-Pack 8$\times$8$\times$8 \textbf{k}-points grids, SCF convergence threshold of $1.36\cdot10^{-7}$ eV and all the other parameters as reported in Electronic Supplementary Information (Table \ref{tab:calc_param}). The PESs are calculated at the resolution of 60$\times$60 single point calculations on shifted structures, spanning the $x$ and $y$ directions from -0.5 Å to 0.5 Å, originating in the five symmetrical points of Figure \ref{fig:stackings}.

\section{Results and discussion}
\label{sec:ResDisc}
\subsection{Stability of the different conformers}
\label{subsec:PES}

\begin{figure*}[!htb]
\centering
 \textbf{}\includegraphics[width=\textwidth]{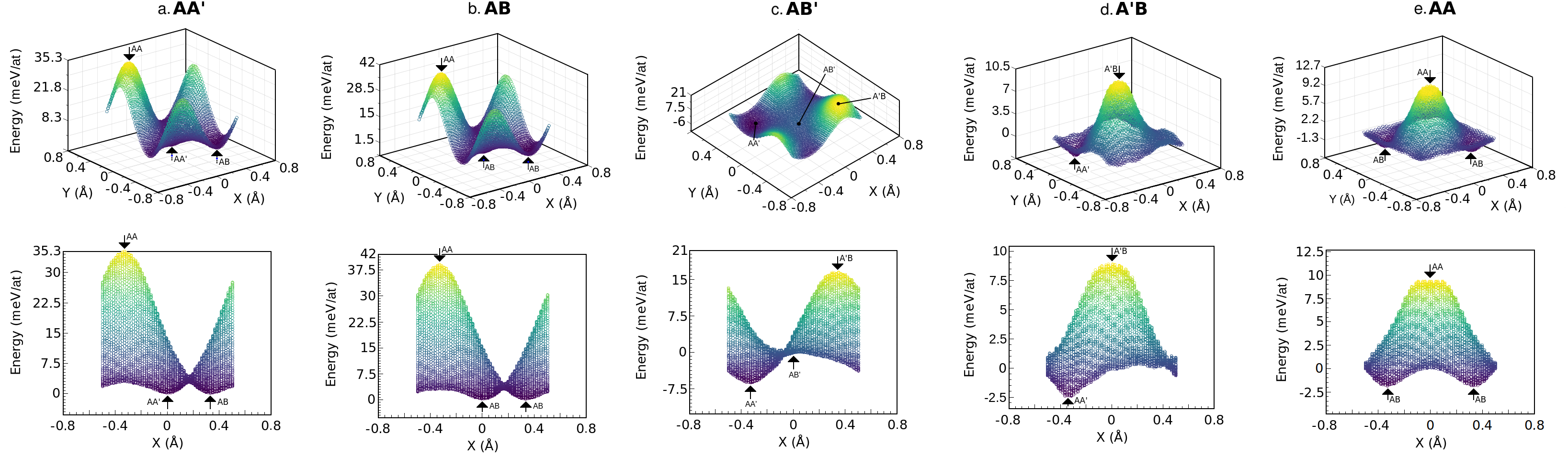} 
\caption{Potential energy surfaces (PESs) cuts originating from the symmetrical points of all five stackings. The cell parameters have been kept constant (sourced from a full structural optimization). The PESs are calculated at the QE-PAW-PBE-D2 level of theory, by sliding of contiguous h-BN planes. Zero points in the energy scales correspond to the energies of the symmetrical structures.}
 \label{fig:PESstable}
\end{figure*}

In Figures \ref{fig:PESstable} and in the Electronic Supplementary Information (Figure \ref{ABprime_PES}) we report potential energy surfaces (PESs) cuts calculated within a selection of different methods. The figures are generated by keeping the $a$ and $c$ lattice constants fixed, as obtained for the five symmetrical structures (reported in Electronic Supplementary Information [Table \ref{cell_par}]) and by displacing the atoms in the lattices with parallel sliding of contiguous h-BN planes with respect each to the other. 
A detailed description of these results is given in the Electronic Supplementary Information. 
The shapes of the PESs confirm the stability of the AA’ and AB symmetrical structures, which is backed up by the comparison of the total energies in Table \ref{tab:energy}. 
\begin{table}[htb]
\centering
  \caption{Relative energies (meV/at) of the various h-BN stacking configurations, with respect to the most stable configuration, calculated for every adopted method.}
  \label{tab:energy}
\scriptsize{ 
\setlength\tabcolsep{4.0pt}
  \begin{tabular}{cllllll}
  \hline
\multicolumn{2}{c}{} & AA'  & AB & AB'   & A'B  & AA  \\
& Method &  $P6_{3}/mmc$ & $P3m1$ & $P6_{3}/mmc$ & $P6_{3}/mmc$ &  $P\overline{6}m2$ \\
\hline
\multirow{4}{*}{\rotatebox[origin=c]{90}{VASP}} & PBE-D2$^{1}$  &  0.27  & 0.00 & 2.34& 15.60 & 18.48 \\
& PBE-D3BJ$^{1}$       & 0.32 & 0.00 & 2.04& 15.63 & 18.29 \\
& PBE-TS$^{1}$         & 2.80 & 0.76 & 0.00& 15.98 & 17.72 \\
& SCAN+rVV10$^{1}$     & 0.42 & 0.00 & 2.49& 17.86 & 21.02 \\
\hline
\multirow{3}{*}{\rotatebox[origin=c]{90}{QE}}
& PBE-D2$^{1}$          & 1.50 & 0.00& 5.37 & 22.87 & 25.22 \\
& LDA$^{2}$              & 0.61 & 0.00& 2.19 & 12.09 & 13.18 \\
& PBE-TS$^{3}$           & 0.22 & 0.00& 1.81 &  9.26 &  9.96 \\
& SCAN$^{3}$             &  1.04    &  0.00 &  4.64 &  24.45  & 26.61 \\
\hline
  \end{tabular}
  \begin{flushleft}
$^{1}$ - Projector Augmented-Wave PP\\
$^{2}$ - Ultrasoft PP, \\
$^{3}$ - Norm-conserving PP.
  \end{flushleft}
 }
\end{table}
Furthermore, the AB' point is located in a prominently flat valley (or a soft groove in NC-SCAN and US-LDA). Although the AA' configuration is easily reachable upon simple sliding, the AB' stacking configuration seems to be metastable, leading to possible interpretations of the experimental data related to a dynamical stability \cite{xu2019optomechanical,henck2017stacking}. The metastability of AB' is particularly evident when examining the shape of the NC-SCAN PES reported in the Electronic Supplementary Information (Figure \ref{ABprime_PES}, uppermost panel).

\begin{figure}[tb]
\centering
  \includegraphics[width=7.5cm]{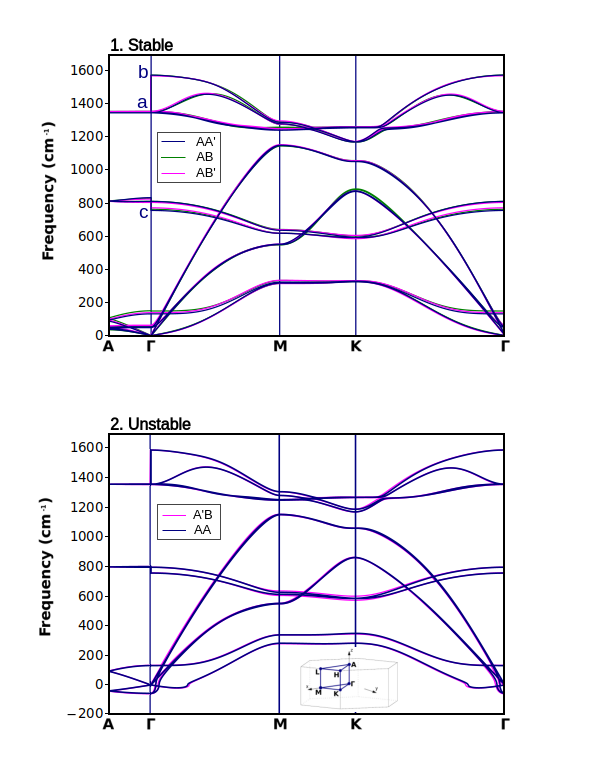}
  \caption{Comparison of the phonon energy dispersions along high symmetry directions in the first Brillouin zone for five h-BN bulk systems.\cite{rem_phonons} IR or Raman active points are indicated as \textit{a, b} and \textit{c} and corresponding displacements are depicted in Fig. \ref{fig:abc_phonon}.}
  \label{fig:phonon}
\end{figure}
\begin{figure}[htb]
\centering
  \includegraphics[height=2cm]{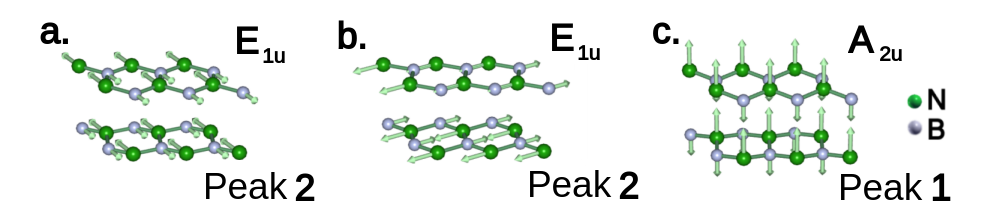}
  \caption{\textit{a}, \textit{b} and \textit{c}: graphical representations of the displacement eigenvectors (green vectors) in the AA' stacked h-BN for three IR or Raman active modes at the $\Gamma$ point (respectively indicated in the Figure \ref{fig:phonon}).}
  \label{fig:abc_phonon}
\end{figure}

We calculated the energetic dispersion of the vibrational modes for the five structures. We employed a wide selection of methods and we provide an extensive description of our results in the Electronic Supplementary Information. While the $\Gamma$ point frequencies of the AA' variant are reported in Table \ref{tab:comp_exp_phon}, we report in the Electronic Supplementary Information all the data for the two other stable structures (Tables \ref{tab:comp_exp_phon_AB} and \ref{tab:comp_exp_phon_ABprime}). A huge effort in interpretation is necessary, since, the nature of the systems and the elusive structure of the physical heterogeneity, the results obtained with different methods often lead to contrasting conclusions. Based on our analysis (see Section \ref{subsec:AccuracyModel}), we selected the most significant dispersion schemes and we report them in Fig. \ref{fig:phonon}, confirming the stability of the stacking structures AA', AB and AB'. 

\begin{table*}[htb]
\centering
  \caption{Phonon frequencies at the $\Gamma$ point (in $cm^{-1}$) for the AA' stacking of h-BN as calculated by different theoretical approaches and software implementations and compared with experimental values taken from literature. The same method used to calculate the IFC was used to obtain the NA except where it is specifically stated otherwise. The same theoretical data are provided for the AB and AB' stackings in the Electronic Supplementary Information (Tables \ref{tab:comp_exp_phon_AB} and \ref{tab:comp_exp_phon_ABprime}).}
  \label{tab:comp_exp_phon}
\setlength\tabcolsep{4.5pt}
\scriptsize{
\begin{tabular}{l|rrrrr|rrrrr|c}
\hline
     & \multicolumn{5}{c|}{VASP}         & \multicolumn{5}{c|}{QE} & \multirow{2}{*}{Experiment} \\
Mode & PBE$^{1}$   & PBE-D2$^{1}$ ($b$)   & PBE-D3BJ$^{1}$ & PBE-TS$^{1}$   & \begin{tabular}[c]{@{}c@{}}SCAN\\ +rVV10\end{tabular}$^{1}$ & LDA$^{2}$ & PBE-D2$^{1}$ ($a$) & PBE-TS$^{3}$ &    \begin{tabular}[c]{@{}c@{}}PBE-TS$^{3}$\\NA: PAW-PBE$^{1}$\end{tabular}  &  SCAN$^{3}$                           \\
\hline
E$_{2g}$**  & 47   & 47   & 46   & 37   & 47   & \begin{tabular}[c]{@{}c@{}}50\\ 51\end{tabular}     & \begin{tabular}[c]{@{}c@{}}60\\ 63\end{tabular}   & \begin{tabular}[c]{@{}c@{}}48\\ 53\end{tabular}   & \begin{tabular}[c]{@{}c@{}}48\\ 53\end{tabular} & \begin{tabular}[c]{@{}c@{}}51\\ 51\end{tabular}  &  51  \cite{Nemanich1981}                        \\
B$_{1g}$  & 126  & 88   & 119  & 138  & 129  & 113    & 182   & 131 &  131   & 113 & -                           \\
A$_{2u}$*  & 745  & 742  & 744  & 744  & 741  & 751    & 722   & 756 & 756  &  736 & 767-810 \cite{Geick1966,hidalgo2013high,ccamurlu2016modification,mukheem2019boron,chen2017thermal,wang2003synthesis,andujar1998plasma}                    \\
B$_{1g}$  & 797  & 793  & 797  & 800  & 795  & 811    & 791   & 810  & 810   & 797 & -                           \\
E$_{2g}$**  & 1354 & 1359 & 1354 & 1356 & 1381 & \begin{tabular}[c]{@{}c@{}}  1384\\ 1384\end{tabular}   & \begin{tabular}[c]{@{}c@{}}1350  \\ 1351\end{tabular}  & \begin{tabular}[c]{@{}c@{}} 1345 \\ 1345\end{tabular}  & \begin{tabular}[c]{@{}c@{}} 1345 \\ 1345\end{tabular}  & \begin{tabular}[c]{@{}c@{}} 1367 \\1367 \end{tabular}  & 1369-1376 \cite{Geick1966,Nemanich1981,Reich2005}                       \\
E$_{1u}$*  & 1354 & 1359 & 1354 & 1356 & 1381 & 1384   & 1351  & 1345  & 1345   & 1367  & 1338-1404 \cite{Geick1966,hidalgo2013high,ccamurlu2016modification,mukheem2019boron,chen2017thermal,wang2003synthesis,andujar1998plasma}                       \\
E$_{1u}$*  & 1589 & 1593 & 1589 & 1591 & 1629 & 1615   & 1591  & 1570  &    1573   &1596  &    1616   \cite{Geick1966}                     \\
\hline
\multicolumn{12}{l}{\begin{tabular}[c]{@{}l@{}}$^{1}$ - Projector Augmented-Wave PP, $^{2}$ - Ultrasoft PP, $^{3}$ - Norm-conserving PP\\ * - IR active modes \\ ** - Raman active modes\end{tabular}}
\end{tabular}
}
\end{table*}

\subsection{Vibrational structure}
\label{subsec:VibSpec_Results}

 Two vibrational modes are IR active in h-BN (Figure \ref{fig:abc_phonon}, described by the irreducible representations $A_{2u}$ and $E_{1u}$), i.e. have $I^{IR}_{\nu} \ne 0$. The ratio between these two intensities can be informative about the microscopical nature of the system \cite{harrison2019quantification,amin2020boron} and it is thoroughly studied here. 
Only one vibrational mode is active in Raman spectroscopy ($E_{2g}$). The frequency of vibration of this Raman active mode is degenerate with that of the IR active $E_{1u}$ mode. Therefore, here we do not investigate Raman tensors.

\begin{figure}[ht]
\centering
   \includegraphics[width=7.0cm]{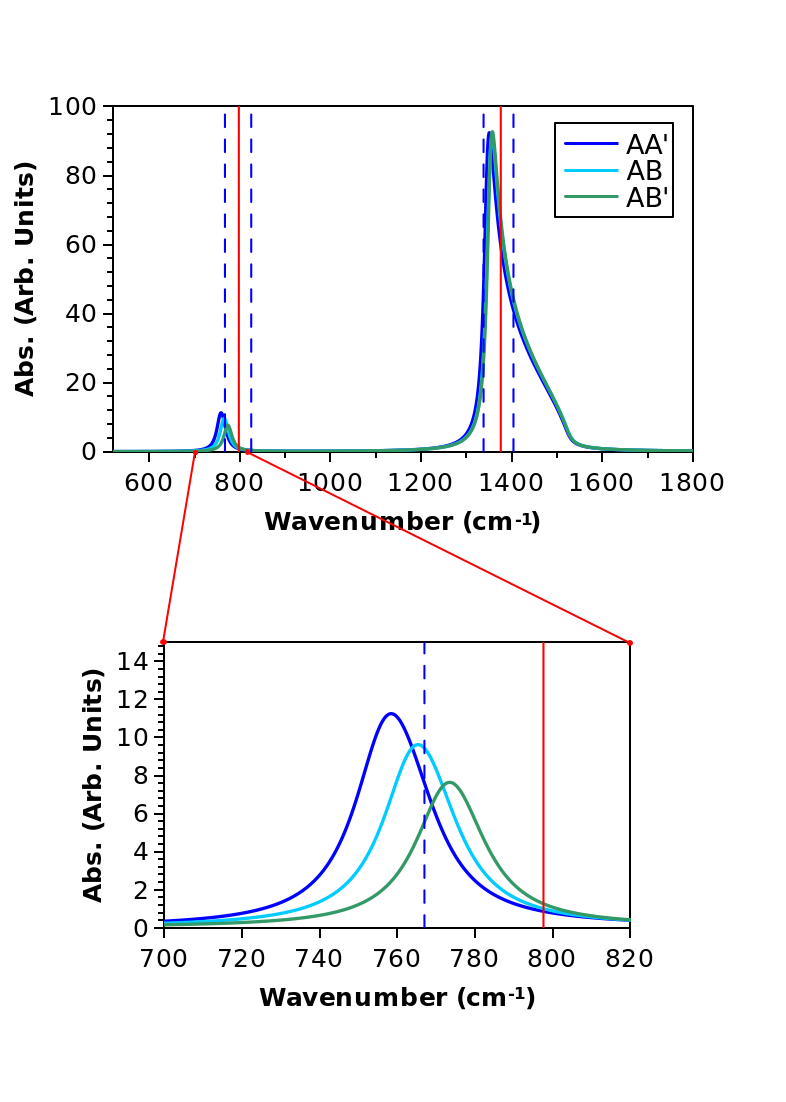} 
  \caption{Calculated vibrational spectra for the three stable h-BN stacking configurations using NC-PBE-TS IFCs and NA parts numerically obtained adopting DFPT at QE-PAW-PBE level. Red solid vertical lines indicate the average experimental peak values among the six cited ones in Table \ref{tab:exp} and blue vertical dotted lines indicate the experimental range (Table \ref{tab:exp}).}
  \label{fig:IR_PBE-D2-TS}
\end{figure}

In Figure \ref{fig:IR_PBE-D2-TS} we report the most significant (based on our analysis, see Section \ref{subsec:AccuracyModel}) vibrational spectra for the three stable stacking structures. The peak frequencies of the AB' spectrum (1396 cm$^{-1}$ and 751 cm$^{-1}$) have higher values with respect to the two other stable stackings: (1355 cm$^{-1}$, 724 cm$^{-1}$ in AA' and 1357 cm$^{-1}$, 733cm$^{-1}$ in AB). The ratio between the intensities of the two active absorptions is 12.129 in the AB' structure which is considerably higher with respect to the other two systems (8.211 in AA' and 9.500 in AB).

\begin{table}[!ht]
\small
\centering
    \caption{Overview of results taken from experimental works concerned with IR spectroscopy in bulk and flaked h-BN samples. In the third column we report the ratio between the two intensities of the main IR absorption peaks as extrapolated from graphical data. In the fourth and fifth column, the frequencies of the two IR active peaks.}
    \label{tab:exp}
    \begin{tabular}{lccccccc}
    \hline
    Label& Source & Ratio&Freq. 1 ($cm^{-1}$) & Freq. 2 ($cm^{-1}$)\\
    \hline
    Sample 1 & Ref. \cite{hidalgo2013high}&   1.805 & 802 & 1365 \\
    Sample 2 & Ref. \cite{ccamurlu2016modification}& 2.636 & 805 & 1381 \\
    Sample 3 & Ref. \cite{mukheem2019boron}& 1.043& 767& 1338\\
    Sample 4 & Ref. \cite{chen2017thermal}& 1.282& 825 & 1378\\
     Sample 5 & Ref. \cite{wang2003synthesis}&1.236&810
&1390\\
      Sample 6 & Ref. \cite{andujar1998plasma}&1.737&777& 1404\\
    \hline
    \end{tabular}
\end{table}

\begin{figure}[htb]
\centering
  \includegraphics[width=7.0cm]{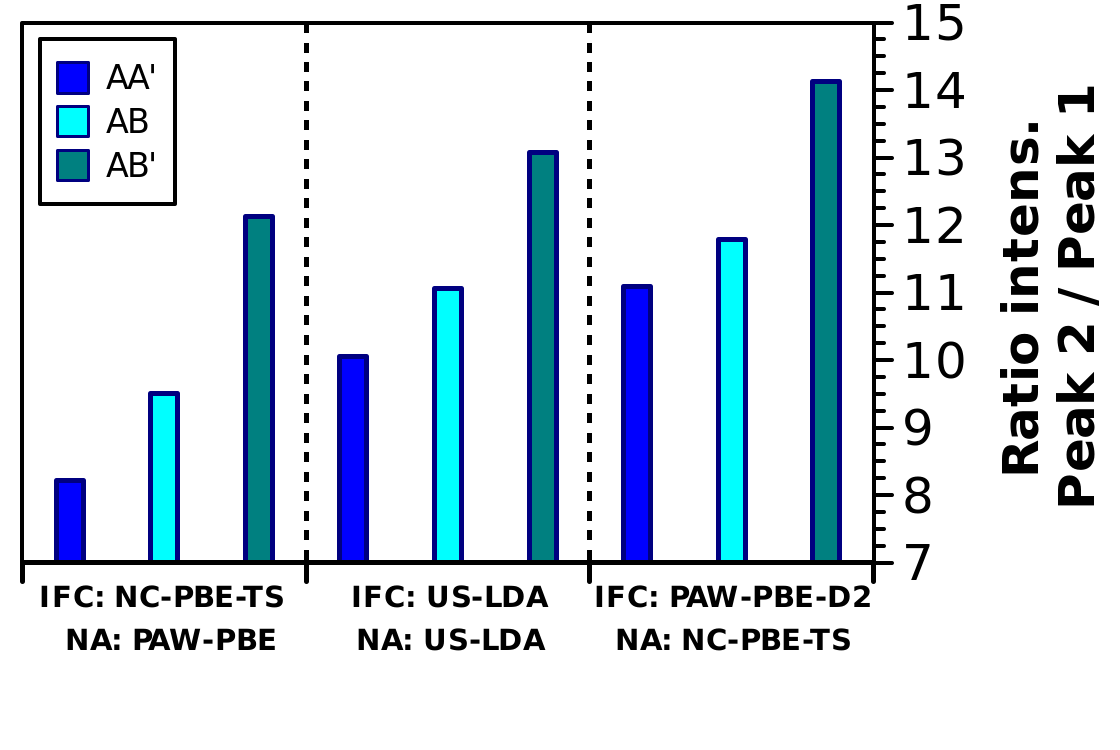}
  \caption{Calculated ratios between the absorption intensities of the two IR active peaks in vibrational spectra of the three stable h-BN stacking structures (AA', AB and AB') by means of three different theoretical implementations (for PAW-PBE we use Set $a$, Table \ref{tab:calc_param} of the Electronic Supplementary Information).}
  \label{fig:fig:IR_theor_ratio_1}
\end{figure}

In Figure \ref{fig:fig:IR_theor_ratio_1} we extend our view and compare the ratios obtained with three different theoretical implementations. The same trend and peculiar behavior ($Ratio_{AB'}>>Ratio_{AB}>Ratio_{AA'}$) is noticeable within all of the tested physical models. In Appendix we show how these numerical results are essentially related to the geometrical structure. 
It is possible to compare the resulting absorption functions with experimental infrared spectra. To date there is no simple way to unravel the stacking composition of samples (we are presenting it in this study) available in literature. For this reason, we have gathered a number of different experimental measurements without insights into  their conformational composition. This collection of data permits us to give an assessment of the quality of the different calculations and, in view of our theoretical results (Section \ref{subsec:EmpDeriv}, Appendix), it works as a testing set for our semi-empirical method (see Section \ref{subsec:EmpDeriv}).

\begin{figure*}[!ht]
\centering
  \includegraphics[height=4.5cm]{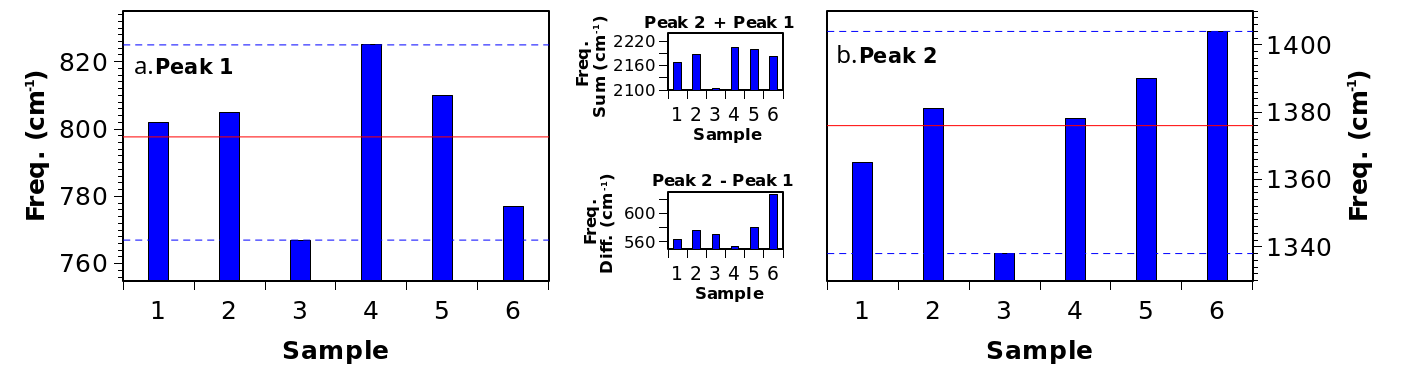}
  \caption{Experimental peak frequencies from the infrared absorption spectra of bulk and flaked h-BN samples analyzed in the text (references in Table \ref{tab:exp}). Red horizontal lines indicate the average values. Blue dotted lines enclose the reported experimental range. In the central insets we compare the summed values and differences.}
  \label{fig:IR_exp_freqs}
\end{figure*}

\begin{figure}[!ht]
\centering
  \includegraphics[height=13.0cm]{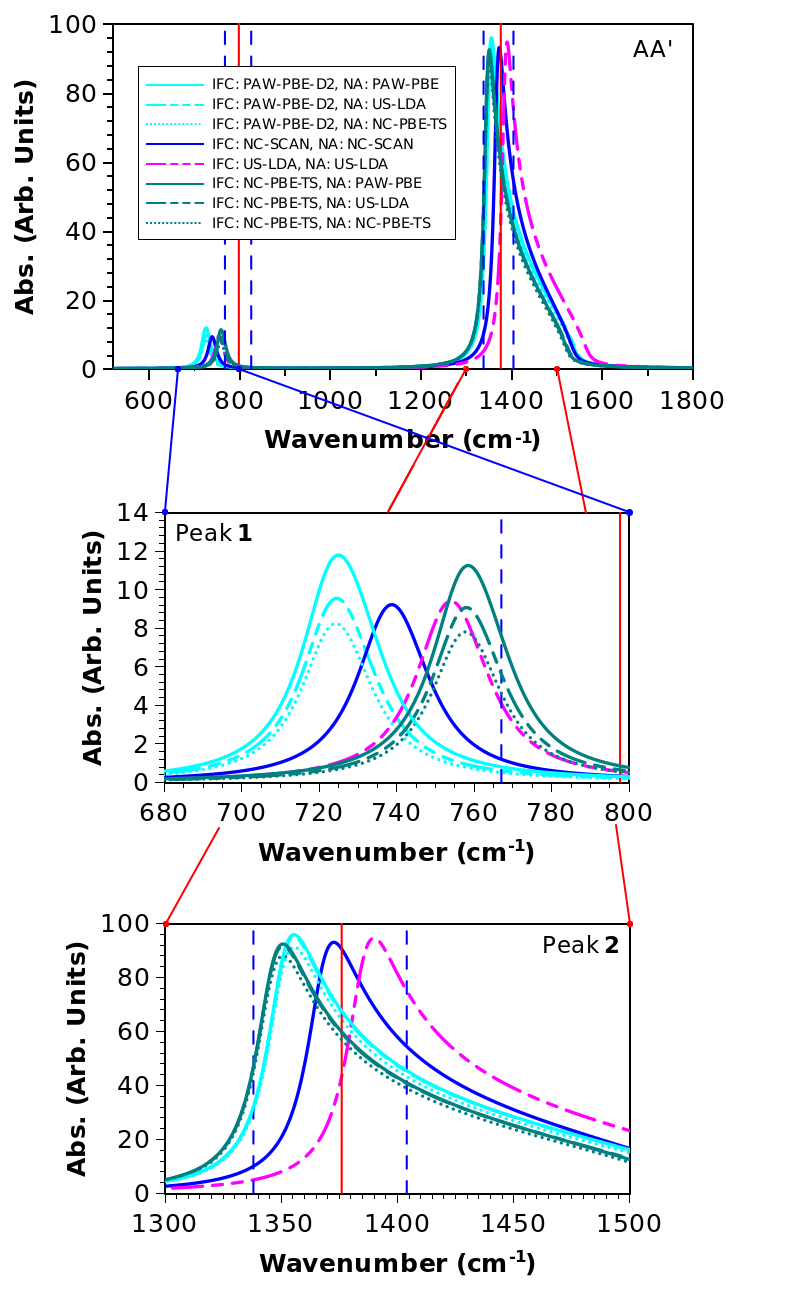}
  \caption{Vibrational spectra calculated for the AA' stacking configuration of h-BN with different theoretical approaches, as described in legend (for PAW-PBE we use Set $a$, Table \ref{tab:calc_param} of the Electronic Supplementary Information). The details of Peak 1 and Peak 2 regions are showed in insets. Red solid vertical lines indicate the average experimental peak values and blue vertical dotted lines indicate the experimental range limits (among the six spectra in Table \ref{tab:exp}).}
  \label{fig:IR_theor_2}
\end{figure}

In Figure \ref{fig:IR_exp_freqs} and in Table \ref{tab:exp}, we report the peak frequencies of the IR experimental set. For comparison with theoretical results we also report the ranges and average values in Figures \ref{fig:IR_PBE-D2-TS} and \ref{fig:IR_theor_2}. A good agreement of the theoretical and experimental data is noticeable in the values of the larger peak (Peak 2). Concerning the second most intense peak (Peak 1), the frequencies exhibit, instead, systematically lower values with respect to the experimental range. 
In general, the experimental frequencies show a wide variability (from 767 to 825 cm $^{-1}$ for Peak 1 and from 1338 to 1404 cm $^{-1}$ for Peak 2). Significant trends can be seen among the samples such as sample number 3 presents the lowest values of both peaks. Comparing the two active peaks, among Samples number 1,2 and 3 the whole spectrum shifts homogeneously while Samples 4 and 6 show a different and opposite order. The anomaly of Samples 4 and 6 is more noticeable when comparing the two central insets in Figure \ref{fig:IR_exp_freqs}. The distance between the two peaks in different samples is reported in the lower inset, while the sum of the frequencies of the two peaks is reported in the upper inset. This different behaviour of Samples 4 and 6 could arise from a number of factors involving systematical structural deformations deeply affecting the BN planar meshes, e.g. ambient temperature, impurities, other effect produced by nanostructuring processes, the description of which goes beyond the purpose of this work.
 
\begin{figure}[!ht]
\centering
  \includegraphics[width=7.0cm]{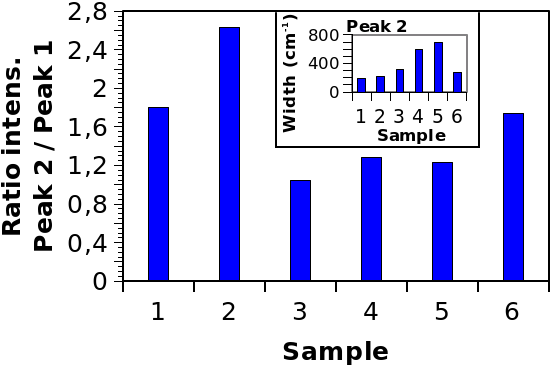}
  \caption{Experimental ratios between the two main IR absorption intensities in six different experimental measurements performed on bulk and flaked h-BN, taken from literature and published by independent research groups (Table \ref{tab:exp}). In inset: for the same experimental spectra (Table \ref{tab:exp}), comparison among the full width at half maximum (FWHM) of the main absorption peak (Peak 2).}
  \label{fig:IR_exp_ratios}
\end{figure}

In Figure \ref{fig:IR_exp_ratios} we report the ratios between the intensities of the two peaks in the referred experimental works. The reported data show a considerable variability, spanning from 2.64 of Sample 2 to 1.04 of Sample 3. 
Considering the ratios and not just the absolute values enabled us to obtain direct information within an internal standard mechanism \cite{harrison2019quantification}. The variability in this parameter is related to structural differences at the atomistic level. Emphasizing this consideration is the complete lack of correlation between the data reported in Figure \ref{fig:IR_exp_ratios} and the width of the experimental peaks measured at 1/2 of the peak height (reported in the inset of the same Figure \ref{fig:IR_exp_ratios}), which instead is affected by the macroscopic features of the specific measured materials (e.g. the grain size, which affect the quasiparticle lifetime by being inversely proportional to the probability of surface scattering events) \cite{lee2012large,henck2017direct,amin2020boron}.

The recent work by Amin \textit{et al.} \cite{amin2020boron} enforces our conclusions, where a first analysis of their results suggests that the ratio between the IR intensities is not due to chemical impurities.

As mentioned before, with the exclusion of the anomalous Samples 4 and 6, the relation between peak values ratios (Figure \ref{fig:IR_exp_ratios}) and peak frequencies (Figure \ref{fig:IR_exp_freqs}) is obvious. Samples 2 and 3 show the highest variability. We reasonably hypothesize that the experimental variability (Figures \ref{fig:IR_exp_freqs} and \ref{fig:IR_exp_ratios}) is due to different conformational composition, including the stacking variants, of the referred samples. Following this hypothesis, we deduce that compared with our theoretical results reported in Figure \ref{fig:IR_PBE-D2-TS}, that Sample 2 contains the highest amount of AB' stacked material in our experimental set, while Sample 3 contains the lowest amount of it.

\subsection{Accuracy of the model}
\label{subsec:AccuracyModel}
In Figure \ref{fig:IR_theor_2} we report the results of calculated vibrational spectra obtained with different theoretical implementations for the AA' stacked system. For an easy comparison, the experimental averaged values and range limit lines are showed in the same graphical scheme. Analogous data, calculated for the other two stable variants are reported in the Electronic Supplementary Information (Figures \ref{IR_theor_2_AB} and \ref{IR_theor_2_AB'}). The agreement between experimental and theoretical values is particularly satisfactory regarding the most intense peak ($E_{1u}$ mode, Peak 2), among all of the considered theoretical perspectives.

The calculated vibrational frequencies of the $A_{2u}$ mode (Peak 1) lay below the lower limit of the experimental range of variability for all employed methods. The presented models underestimate the energy necessary for the $A_{2u}$ vibrational movement (see Figure \ref{fig:a2u}). 
As seen in Figure \ref{fig:IR_theor_2}, changing only the NA part of the dynamical matrix does not produce significant changes in the peak frequencies. The differences among the results obtained in calculated frequencies are mainly attributable to the IFC part. 

There are several aspects of the vibrational spectra that indicate the accuracy of the employed method. The first criterion to assess the accuracy is the distance of frequencies of vibration for the $A_{2u}$ mode (corresponding to Peak 1) from the lower edge of the experimental range(See the Peak 1 enlargement, in the lower part of Figure \ref{fig:IR_theor_2}). In regard to this, the best performance is achieved by the NC-PBE-TS IFC, while the worst results are produced by the PAW-PBE-D2 IFC. A second assessment criterion are the computed intensities of the two IR active peaks. In particular one should considered the ratio between the  values of the two peaks \cite{harrison2019quantification}, as discussed in Section \ref{subsec:VibSpec_Results}.

\begin{figure}[htb]
\centering
  \includegraphics[width=6.8cm]{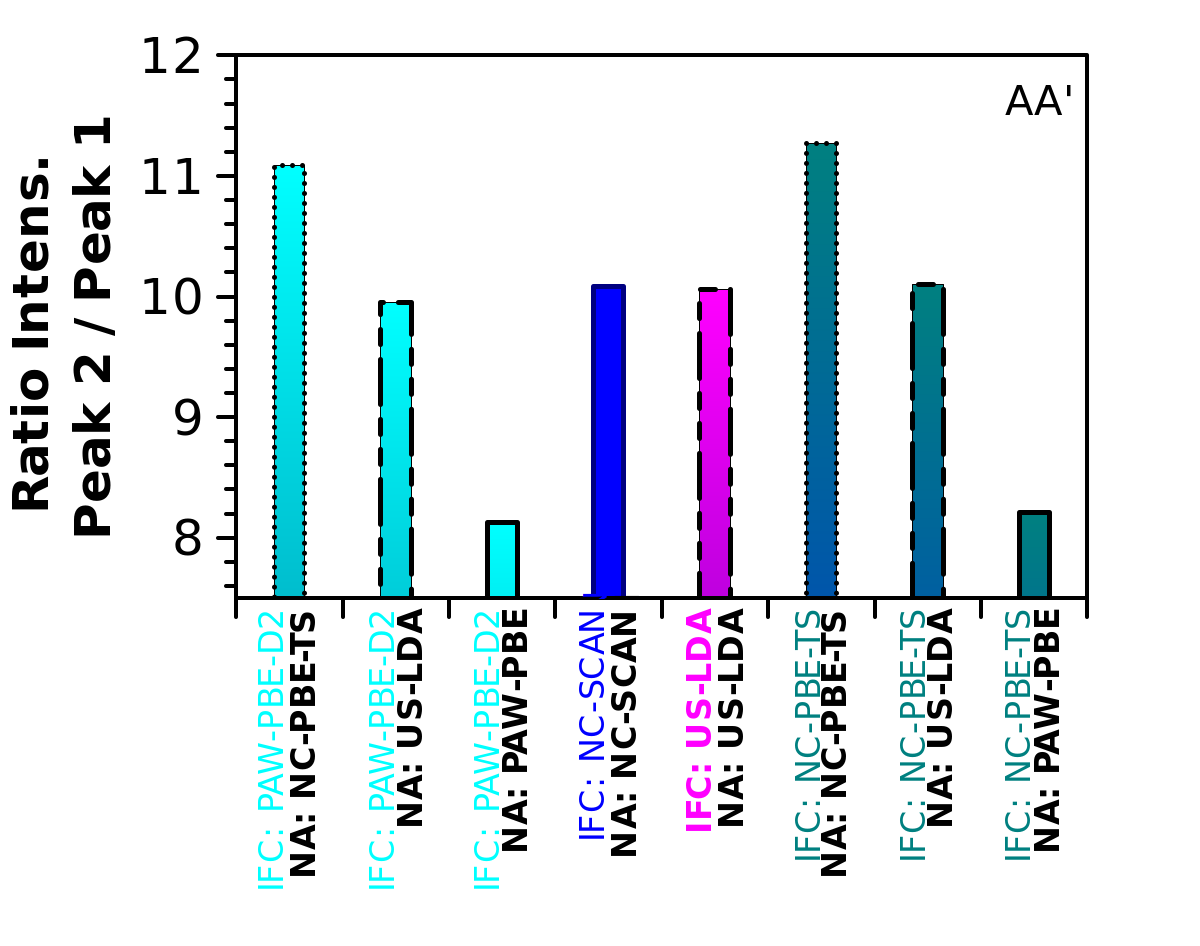}
  \caption{Calculated ratios between the two IR active absorption intensities in vibrational spectra obtained for the simulated AA' stacking structure of bulk h-BN with different theoretical implementations (for PAW-PBE we use Set $a$, Table \ref{tab:calc_param} of the Electronic Supplementary Information).}
  \label{fig:IR_theor_ratio}
\end{figure}

The gathered experimental values of peak ratios (Figure \ref{fig:IR_exp_ratios}, Table \ref{tab:exp}) span a considerably narrower interval and have a lower average with respect to the calculated ones. E.g. with QE-PAW-PBE-D2, IFC the theoretical range goes from 8.13 (with QE-PAW-PBE NA part) to 11.27 (with NC-PBE-TS NA part). In Figure \ref{fig:IR_theor_ratio}, we report a wide collection of calculated ratios, using different theoretical approaches for both IFC and NA parts. From the comparison, it is clear that the main source of variability is the NA part of the dynamical matrix.
Analyzing the two inserts of Figure \ref{fig:IR_theor_2}, we notice that the variability in the value of IR active peak ratios, mainly derives from the intensities of Peak 1.

A general consideration of the two assessment criteria shows a substantial failure of the employed methods (DFT with the common dispersion corrections) in the description of the $A_{2u}$ vibrational mode physics. Since the dynamics of this mode reflect predominately the stacking interactions in h-BN, this failure indicates the unsuitability of the model for use in the description of stacking dynamics of our system.

It seems that one method can not correctly describe both the IFC as well as NA part of the dynamical matrix. This can be shown for the NC-PBE-TS, which delivers the best performance in the calculation of the IFC, while being the worst method in the calculating the NA, and vice versa, for the PAW-PBE-D2 method. 

This is to be expected since it is well known that the GGA methods describe the covalent bonds in a superior way with respect to the LDA methods. We show that LDA exhibits a surprisingly good performances in the calculation of phonon frequencies and an average performance regarding the reciprocal intensities of absorption peaks which has also been observed in previous works \cite{cusco2018,mosuang2002relative,janotti2001first,hamdi2010ab}. US-LDA results to be a theory of intermediate quality for both assessment criteria. The best results overall were obtained using a complex approach implementing GGA (IFC: NC-PBE-TS, NA: QE-PAW-PBE).

In Appendix we rigorously show, in an original manner, how the results obtained by varying the theoretical approaches can give dynamical information about the process of charge density fluctuation related to the $A_{2u}$ vibrational mode. Following this consideration, we can assert that the effective charge density resulting from the US-LDA calculation (restricting the assertion only to the nitrogen hypothetical polar cones, see Appendix), is more similar to the "apparent charge density" $\rho_{(Q)}(r)$ of the real system than the one produced by NC-PBE-TS but less similar to the PAW-PBE effective charge density. In this regard LDA is effectively considered a well-performing theory. On the other hand, the good performances of LDA also depends on its poor description of the covalent bonds: the underestimation of the bond strengths results  in an underestimation of the spring recall forces which compensate the inadequacy of the models to account for the van der Waals stacking interaction.

\subsection{Semi-empirical calculation of the dielectric dynamical parameters}
\label{subsec:EmpDeriv}

\begin{figure}[!ht]
\centering
  \includegraphics[width=6.6cm]{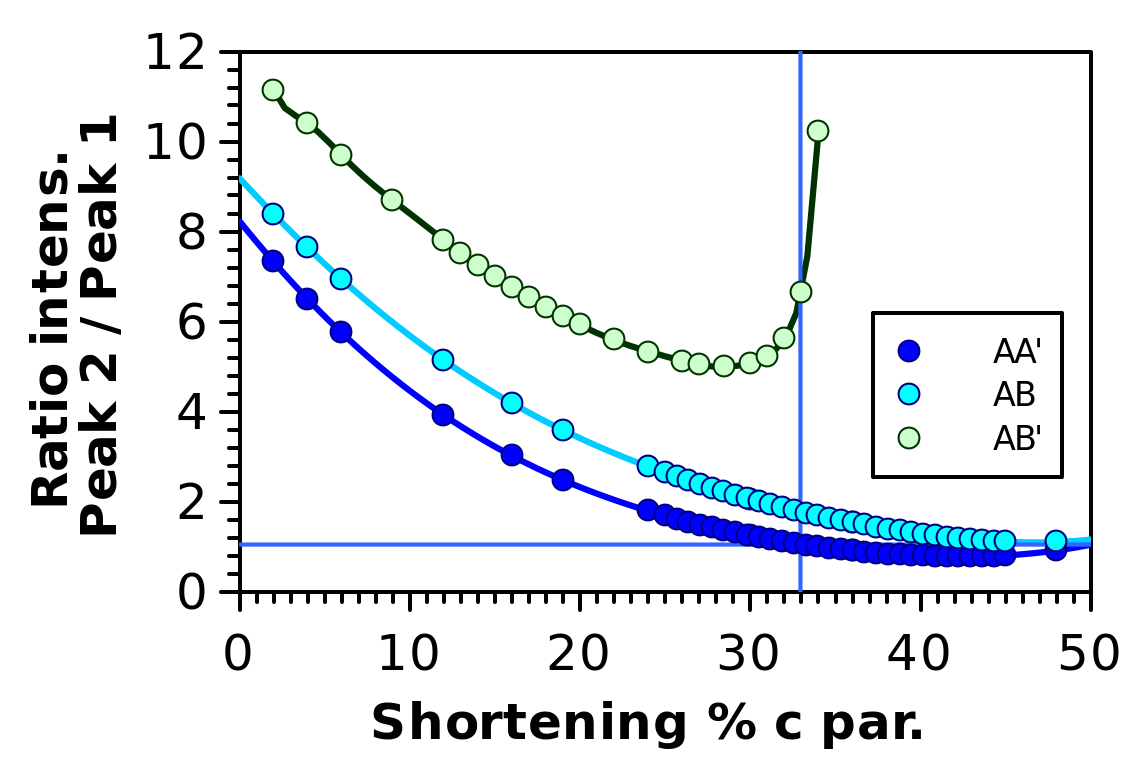} 
  \caption{Calculated ratio between the two IR active absorption intensities for the three stable stacking structures of h-BN reported as functions of the shortening percentage applied to the $c$ structural parameters. Each point results from a numerical calculation of the relevant NA parameters and IR absorption spectral function. The continuous interpolation lines are calculated by fitting second degree polynomial expressions for AA' and AB and a twelfth degree polynomial for AB'. The horizontal (blue) line is drawn at 1.043 (Sample 3 of the experimental set). The vertical (blue) line (32.94\%) intercepts the horizontal one and the AA' interpolation line.}
  \label{fig:sh_cell}
\end{figure}

The results reported in Figure \ref{fig:sh_cell} are obtained on fictitious systems, which are produced by arbitrary shortening of the $c$ structural parameter from the fully optimized structures QE-PAW-PBE-D2 (cell parameters in Electronic Supplementary Information [Table \ref{cell_par}]), without any further geometrical optimization. The figure displays the ratio, calculated between the IR absorption peaks, as a function of the shortening percentage. The data are obtained employing the best performing methods among the tested ones, as discussed in the previous section. The Lorentz broadening functions are always applied separately to each phonon spectral line (Peak 2 is composed by four spectral lines, Peak 1 by one spectral line).

We interpolate the AA' and AB points with parabolic functions and the AB' points with a twelfth degree polynomial function. The AB' function lies systematically above the other two, presenting consistently higher values of ratios. 
The trend of the AB' function exhibits an asymptotic behavior at about 35\%, making it unreasonable to choose any higher number (at least for the specific AB' configuration).

As an educated guess, we use, here, the same value of shortening (in percentage, starting from the respective optimized geometries) for the three stable systems. The choice of the shortening is arbitrary and is presented here as a first attempt (expected to be debated in subsequent contributions) to obtain bits of information about the microscopical stacking composition of real samples in a simple way.
We propose an optimal shortening of 32.94 \% of the $c$ parameter. The decision comes from the consideration of experimental Sample 3 as purely composed by AA' stacked material. The chosen value corresponds to the vertical line drawn at the intersection between the AA' function in Figure \ref{fig:sh_cell} and the Sample 3 horizontal line, in the same figure.

In the Section \ref{subsec:TheorDescr} we showed how the construction of fictitious systems, with a shortened structural parameter perpendicular to plane directions, permits one to obtain a spatial charge density function more similar to the effective function $\rho_{(Q)}(r)$, at least in the nitrogen polar cones $(R_{HP}, \theta_{HP})$. A detailed discussion of these aspects is given in Appendix.

The real space charge density functions $\rho_{(QE-PAW-PBE)}^{fict.}(\textbf{r})$ of these fictitious systems are proposed, here, as the best obtained approximation of $\rho_{(Q)}(\textbf{r})$ (in the described regions of space). Nitrogen $xz$ sections of $\rho_{(QE-PAW-PBE)}^{fict.}(\textbf{r})$ and the resulting NA parameters, calculated by numerical implementation of DFPT QE-PAW-PBE, are reported in Electronic Supplementary Information (Fig. \ref{Rho_emp}, \ref{Rho_detail} and Tab. \ref{epsilon_emp}).

\begin{figure}[!ht]
\centering
  \includegraphics[width=8.9cm]{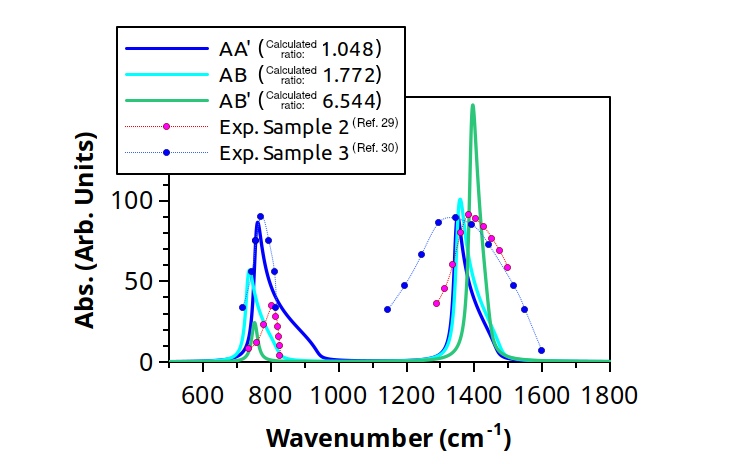} 
  \caption{Vibrational spectra calculated for the three stable stacking structures of h-BN by NC-PBE-TS (for the analytical IFCs) and semi-empirical NA parts of the dynamical matrices obtained with fictitious systems (as explained in Section \ref{subsec:EmpDeriv}). Dotted lines are obtained by graphical extrapolation of selected points from the cited experimental works \cite{mukheem2019boron,ccamurlu2016modification} (for these, the measured percentage of absorbance is given, not arbitrary units) and are reported for comparison purposes.}
  \label{fig:IR_emp}
\end{figure}

The vibrational spectra calculated with these NA parameters and NC-PBE-TS IFC matrices are reported in Fig. \ref{fig:IR_emp}. The calculated ratios ($Rx_{AA'}$, $Rx_{AB}$ and $Rx_{AB'}$) are reported in the inset of the same figure. Note that the Lorentz broadening functions are always applied separately to each phonon spectral line (i.e. Peak 2 is composed by four spectral lines, Peak 1 by only one spectral line). Linear convolutions of these functions can reproduce exactly the experimental lines and the results of this trial calculation are shown in Table \ref{tab:calc_comp}. Note that, to obtain these values, the IR spectra of Fig. \ref{fig:IR_emp} must be shifted in frequency in order to have a correspondence of the peaks. The multiplicative coefficients are reported as calculated fractions ($AA'(\%)$, $AB(\%)$ and $AB'(\%)$) of the three different stacking conformations. Again, we point out that these results are based on reasonable (but arbitrary) hypotheses, nevertheless, they show the ease of access of the stacking information with infrared spectroscopy.

\begin{table}[!ht]
    \centering
    \scriptsize
    \caption{Hypothetical compositions of the cited experimental samples, following the premise that Sample 3 is purely composed by AA' stacked material. With the coefficients in this table, the linear convolution of the three stacking specific spectra reported in Fig. \ref{fig:IR_emp} (shifted in frequency in order to superimpose the two peak values) exactly reproduces the experimental results (reported in Column 3).}
    \label{tab:calc_comp}
\setlength\tabcolsep{3.0pt}
\begin{tabular}{lllllll}
\hline
&Label & Source & Exp. Ratio &  AA' (\%)    & AB (\%)  & AB' (\%)  \\
\hline & &  & &\textbf{HYPOTHESIS}\\
& Sample 3   & Ref. \cite{mukheem2019boron}       & 1.043 & 100.00 & 0.00 & 0.00 \\
\hline & &  & &\textbf{$\implies$CALCULATION}\\
& Sample 1  & Ref. \cite{hidalgo2013high} & 1.805 & 30.74 & 51.00 & 18.27  \\
& Sample 2  & Ref. \cite{ccamurlu2016modification}     & 2.636 & 9.27 & 51.00 & 39.73 \\

& Sample 4   & Ref. \cite{chen2017thermal}  & 1.282 & 59.29 & 40.02 & 0.69 \\
& Sample 5   & Ref.	\cite{wang2003synthesis}        & 1.236 & 66.75 & 32.53 & 0.72  \\
& Sample 6   & Ref.  \cite{andujar1998plasma}          & 1.737 & 30.39 & 55.04 & 14.56  \\
\hline
\end{tabular}
\end{table}



The hypothesis that the remarkably low value of intensity ratio between the two IR peaks in Sample 3 is produced by an almost exclusive  presence of the AA' stacking variant means that, in accordance with our dynamical model, we decide to exclude a purely enthalpic behaviour (namely a constant, almost unitary ratio, between the amounts of AA' and AB variants in all of the samples). An enthalpic description, in fact (e.g. applied to Sample 3: 50\% AA', 50\% AB and absence of AB') would result in a shortening percentage above the asymptote for AB', contrasting one of our assumption (the same shortening percentage should be suitable for the three variants). Enthalpic considerations have, nevertheless, been applied in case of multiple solutions, keeping the fraction of AB' to be the lowest possible.

\section{Conclusions}
\label{sec:Conclusions}
We compared the phononic structures of five possible stacking configurations of bulk h-BN with experimental outcomes (gathered here from literature review). The comparison is based on two distinct criteria: (1) the agreement between the calculated phononic frequencies (of selected vibrational modes, active in infrared or Raman spectroscopy, at $q=\Gamma$) and the experimental counterparts and (2) the evaluation of the ratio between the intensities of the two IR active peaks. 

We provided results for different DFT theoretical implementations, comparing GGA, SCAN and LDA functionals, as well as NC, US and PAW pseudopotential approximations and different treatments of the van der Waals dispersion correction. 
Differently from what previously reported by other authors, we obtained better results from GGA functionals rather than from LDA, nevertheless contradictory conclusions with respect to the investigated PP approximations, finding that the FHI Troullier-Martins NC PP implementations produce a better agreement of the eigenvalues with respect to experimental measures, while Kresse-Joubert PAW PP deliver better performances in approaching the experimental order of magnitude regarding the ratios between IR absorption intensities. 
LDA is presented as a uniform theoretical description, able to sufficiently describe the vibrational properties of bulk h-BN. Here we found, instead, that a complex theoretical approach, based on GGA, better resolves the heterogeneous physics of the examined systems and produces a closer agreement with respect to the experimental numbers. We reported, instead, scarce or no influence on the results from the different formal treatments of the van der Waals dispersion corrections.

The analysis of the PES surfaces produced by parallel shifting of h-BN planes confirms the stability of the AA' and AB structures. Besides to these, the surrounding area nearby the AB' symmetry point presents the features of a wide \textit{plateau} and significant dynamical stability for this configuration can be reasonably advanced. An extensive analysis of the energy dispersion schemes for the phononic modes confirmed these hints for stability.

A dynamical stability of the AB' conformation, variable upon experimental conditions, produces a systematical presence, in different amounts, of this structured material in real samples. This conclusion would theoretically explain the reported wide range of experimental variability for the ratio between the intensities of the two IR active peaks. Our study resulted in a confirmation, never reported before, that the signal from the AB' stacked variant (ratio between the intensities of the two IR active peaks) is clearly distinguishable and informative. Our results have been directly related to (and confirmed by) purely geometrical considerations (Appendix, see Figure \ref{fig:Disp_rho} at a glance), by means of an effective charge density method (called here "apparent charge density", citing Cochran and Cowley \cite{cochran1962dielectric}). We proposed a simple semi-empirical way for the calculation of the non-analytical part of the dynamical matrices (Born $Q$ matrices and dielectric tensors) for layered systems. We showed, presenting a wide amount of theoretical data, that a multifaceted physical approach is necessary and that an effective charge density method could turn out to be the most convenient pathway for it in layered systems. 

In view of our theoretical findings, we invite to consider the possibility that a significant amount of information about the h-BN stacking variability of multilayered real systems could be simply extracted by the exclusive use of infrared spectroscopy and vibrational analysis.  
An early experimental application of it (very recent) can be found in Harrison \textit{et al.} \cite{harrison2019quantification}. Due to the lack of theoretical literature on the matter, the cited authors do not fully depict the problem, which is, instead, clearly explainable accounting the results presented here.

As this is a first theoretical effort, it is clear that more extended studies (stacking-specific [peculiar] experimental data sets, validation methods and theoretical models) are expected, to enable any application of these conclusions.

\section*{Acknowledgements}
\label{sec:Acknowledgements}
This work was supported by
Czech Science Foundation (18-25128S), 
Institution Development Program of the University of Ostrava (IRP201826),  InterAction program (LTAIN19138),
and IT4Innovations National Supercomputing Center (LM2018140).

\section*{Appendix: Derivation of the dielectric dispersion dynamics from geometry}
\label{Appendix}
By the definition of "apparent charge density" given above (Semi-empirical description, Section \ref{subsec:TheorDescr}) and in view of the obtained numerical results reported in Figure \ref{fig:IR_theor_ratio}, considering the calculated DFPT $Q^{* \alpha   \beta}_{s}$ and $\epsilon$ as numerical counterparts of the $Q$ matrix elements and $\epsilon^{(\infty)}$ matrices as defined in the current work (Methods Section \ref{subsec:LatDyn}), in a conical system of coordinates centered on the nitrogen nuclei:

\begin{equation}
\small{
\begin{aligned}
dxdydz= dR   \left(  sin{\phi} tan{\theta} dR +R cos \phi \tan\theta  d\phi + \frac{R sin\phi  d\theta}{cos^{2}\theta}\right) \\
\cdot  \left( -R \sin{\phi} \tan{\theta}  d\phi  + \cos{\phi} \tan{\theta}  dR +\frac{R \cos{\phi}   d\theta }{{\cos^{2}{\theta}}}   \right),
\end{aligned}
}
\end{equation}

it is possible to state, with respect to the nitrogen nuclei, the existence of polar cones of space $(R_{PC}, \theta_{PC})$, characterized by:

\begin{equation}
\begin{split}
    P_{(NC-PBE)}(R_{PC}, \theta_{PC}) \leq  P_{(US-LDA)}(R_{PC}, \theta_{PC})  \\ \leq  P_{(PAW-PBE)}(R_{PC}, \theta_{PC})  \leq  P_{(Q)} (R_{PC}, \theta_{PC})  ,
\end{split}
\end{equation}

where $P_{(x)} (R_{PC}, \theta_{PC})$ are integrated charge densities in the fraction of space included in the polar cones. The indices $(x)$ represent the different theoretical methods used to obtain the charge density functions $\rho_{(x)}(R, \theta, \phi)$. $\theta_{PC}$ is the conical angle, $R_{PC}$ is the cone height along the $z$ axis and $\phi$ is the $xy$ planar angle:

\begin{equation}
\small{
P_{(x)}(R_{PC}, \theta_{PC})=\iint_{R \leq  R_{PC}, \theta \leq \theta_{PC}}  dR d\theta  \int_{0}^{2\pi}  \rho_{(x)}(R, \theta, \phi) d\phi     }  .
\end{equation}

$\rho_{(Q)}(R, \theta, \phi)$ is the  "effective charge density" as given above. 

We plausibly hypothesize the existence of a second type of nitrogen polar cones, which we define by the hypothetical parameters $R_{HP}$ and $\theta_{HP}$, in which:

\begin{equation}
\label{eq:hp_cones}
\begin{split}
    \bigg[ \rho_{(NC-PBE)}(R, \theta,  \phi) \leq  \rho_{(US-LDA)}(R, \theta,  \phi)  \\ \leq  \rho_{(PAW-PBE)}(R, \theta,  \phi)  \leq  \rho_{(Q)}(R, \theta,  \phi)  \bigg]_{R \leq  R_{HP}, \theta \leq \theta_{HP}}.
\end{split}
\end{equation}

Stated the properties of the radial solutions of the Schrödinger equation, $R_{HP}$ can be chosen in order to be:

\begin{equation}
\begin{aligned}
\big[R  > R_{HP}\big]_{ \theta \leq \theta_{HP} }\implies  \bigg[ \frac{\partial \rho_{(NC-PBE)}(R, \theta, \phi)}{\partial R} \\ =       \frac{\partial \rho_{(US-LDA)}(R, \theta, \phi)}{\partial R}    \\   =   \frac{\partial \rho_{(PAW-PBE)}(R, \theta, \phi)}{\partial R} \bigg].
\end{aligned}
\end{equation}

\begin{figure}[htb]
\centering
  \includegraphics[height=17.2cm]{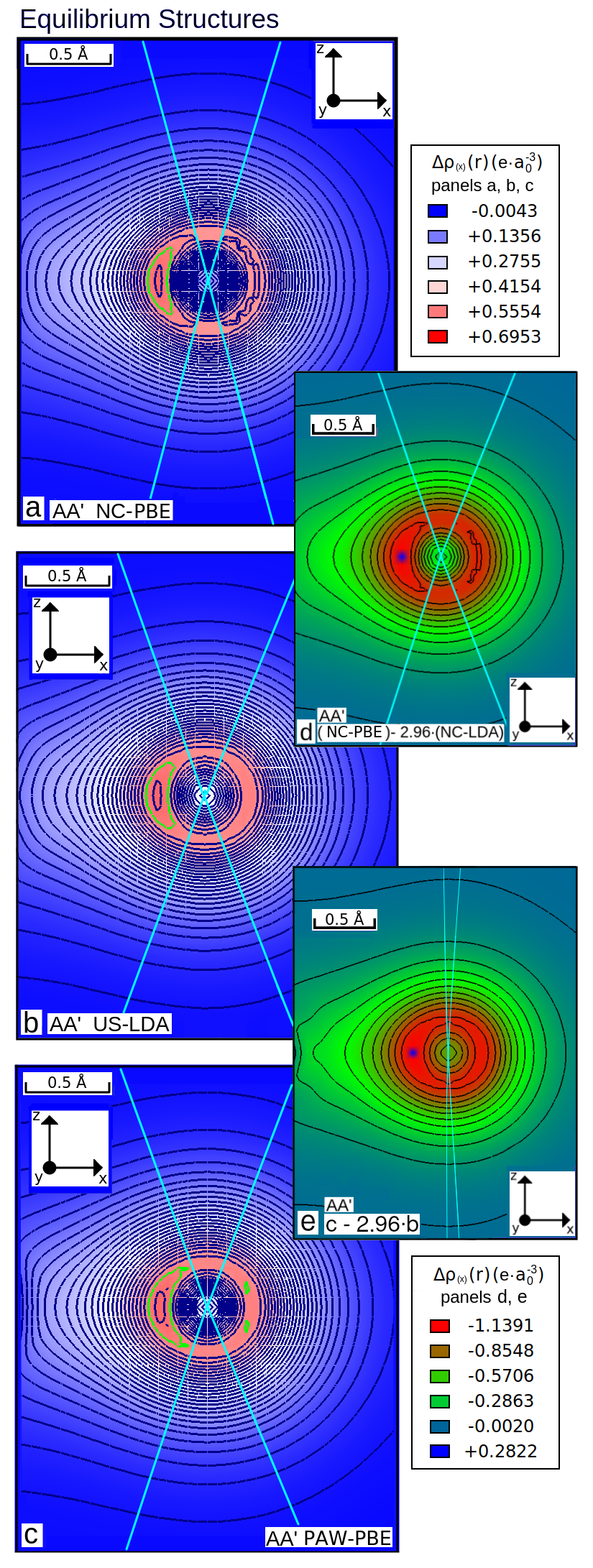}
  \caption{($a$, $b$, $c$) Valence electronic charge density functions (recalculated with a $40\times 30 \times 20$ Monkhorst-Pack \textbf{k}-points grid in the first Brillouin zone of an equivalent orthorhombic unit cell), in the nitrogen $xz$ real plane (atomic centres in the middle of the panels) for the optimized AA' stacked structure, obtained by means of three different theoretical methods: ($a$) NC-PBE, ($b$) US-LDA, ($c$) PAW-PBE (Set $a$). In insets ($d$, $e$): normalized difference functions between charge densities calculated with different methods as reported in the bottom-left of the panels (where a, b and c are the functions depicted in the left panels, NC-LDA: PW cutoff 2041 eV for diagonalization, 8163 eV for the charge density). Light blue lines indicate the projections of polar cones in which it is not possible to identify an angular order in the morphological structure of the isoline systems. In Panels $a$, $b$ and $c$, the 0.5718 $e \cdot a_{0}^{-3}$ isolines of charge density are highlighted in green. Close-up details in Electronic Supplementary Information (Fig. \ref{Rho_detail}).}
  \label{fig:GS_rho}
\end{figure}

In this case $R_{HP}$ is smaller than the atomic radius of nitrogen atom (0.65 {\AA}) and the validity of Eq. \ref{eq:hp_cones} depends only on the choice of $\theta_{HP}$.
A right choice of it implies that the different charge densities reported in Figure \ref{fig:GS_rho}, concerning just the part of space included in hypothetical cones $(R_{HP}, \theta_{HP})$, approximate different phases of the fluctuation process of the charge density, relevantly to the $A_{2u}$ vibrational mode (Peak 1) in purely crystalline, homogeneously stacked real systems.

In Figure \ref{fig:GS_rho} we highlight conical regions in the poles of nitrogen atoms, in which no angular structure is detectable in the reported charge density functions.

An important radial structure can be described (truncated spherical crown, colored in hues of red in all of the panels in Fig. \ref{fig:GS_rho}) extended from 0.2 to 0.4 \AA\ from the nitrogen nuclei. 

Analyzing the difference functions, Panels $d$ and $e$ in Figure \ref{fig:GS_rho}, we notice that the regions colored in red are the same ones, presenting high values of difference. Stated the polarization of the $A_{2u}$ mode along the $z$ axis, it could be easily showed mathematically that the hypothetical cones $(R_{HP}, \theta_{HP})$ are contained in the light blue section lines of Figure \ref{fig:GS_rho}. In this hypothesis (existence of $[R_{HP}, \theta_{HP}]$), the red truncated spherical crowns from 0.2 to 0.4 \AA\ in the polar regions of nitrogen atoms are places of a major charge density variation in the fluctuation movement associated to the $A_{2u}$ vibrational mode (Peak 1).

\begin{figure}[!ht]
\centering
  \includegraphics[height=7.9cm]{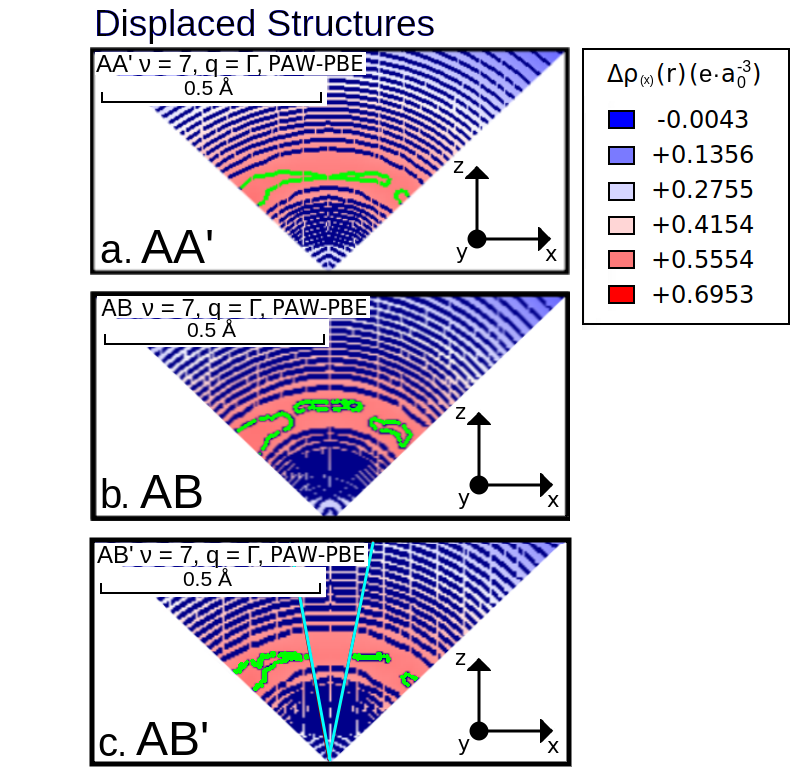}
  \caption{($a$, $b$, $c$) In displaced structures with application of the $A_{2u}$ phononic eigenvectors: valence electronic charge density functions (PAW-PBE [Set $a$], recalculated with a $40\times 30 \times 20$ Monkhorst-Pack \textbf{k}-points grid in the first Brillouin zone of an equivalent orthorhombic unit cell), in the $xz$ real plane sections of polar cones ($\theta=45^\circ$) with vertex in the centres of the nitrogen atoms. The displaced structures are obtained starting from the relevant (PAW-PBE-D2 [Set $a$]) optimized ($a$) AA', ($b$) AB, ($c$) AB' layouts and applying displacement vectors as resulting eigenvectors of the structures of Figure \ref{fig:phonon} ($\nu=7, q=\Gamma$). The 0.5718 $e \cdot a_{0}^{-3}$ isolines of charge density are highlighted in green, being the most meaningful feature detectable at the depicted level of resolution. In Panel $c$, light blue lines indicate the projection of a polar cone in which it is not possible to detect an angular structure in the isoline system. The same characteristic can not be evidenced in the other two structures, as described in the text. Close-up details in Electronic Supplementary Information (Fig. \ref{Rho_detail}).}
  \label{fig:Disp_rho}
\end{figure}

In Figure \ref{fig:Disp_rho}, we report, instead, the electronic charge density functions calculated in displaced structures, i.e. structures in which displacement vectors have been applied to atoms as resulting from phonon eigenvectors ($A_{2u}$ vibration mode: IR Peak 1, $\Gamma$ point). A macroscopic difference can be observed among the functions calculated from different stacking geometries. This difference is well depicted from the 0.5718 $e \cdot a_{0}^{-3}$ isoline in Figure \ref{fig:Disp_rho} (marked in green), and is contained exactly in the highly fluctuating region (red truncated spherical crown from 0.2 to 0.4 \AA\ from nitrogen nuclei) and noticeable in the polar regions of nitrogen atoms (see the light blue conic lines in Panel $c$, Figure \ref{fig:Disp_rho}), relating the higher ratio between the IR active absorptions of the AB' structure (Figure \ref{fig:fig:IR_theor_ratio_1}) to simple geometrical considerations.

\newcommand{\fakesection}[1]{%
  \par\refstepcounter{section}
  \sectionmark{#1}
  \addcontentsline{toc}{section}{Supplemental Material}
}

\fakesection{Supplemental Material}

\begin{figure*}[htb]
\Large
\  \\
\  \\
\  \\
\textbf{Electronic Supplementary Information}\\
\  \\
\  \\
\  \\
\normalsize

\centering
  \includegraphics[height=20cm]{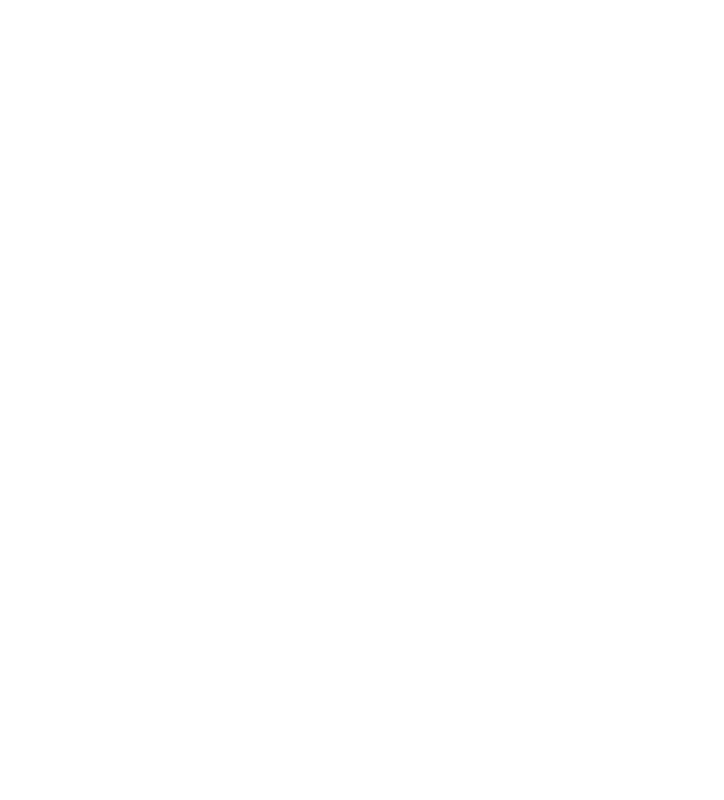}
\end{figure*}

\begin{figure*}[htb]
\Large
\  \\
\  \\
\  \\
\textbf{}\\
\  \\
\  \\
\  \\
\normalsize
\begin{flushleft}
\subsection*{List of provided supplemental data:}
{TS - Supplemental table.}

{S - Supplemental figure.}

\begin{itemize}
    \item \ref{tab:calc_param} - Detailed parameters of all the calculations performed.
    \item \ref{cell_par} - Lattice parameters calculated with a selection of different methods.
    \item \ref{tab:coordinates} - Atomic coordinates of the symmetrical structures.
    \item \ref{ABprime_PES} - PESs cuts for rigid sliding of h-BN planes from the AB' structure calculated with a selection of different methods.
    \item \ref{epsilon_emp} - Semi-empirical Born Q matrices and dielectric tensors calculated as explained in Sections III C and V E of the Main Text.
    \item \ref{tab:comp_exp_phon_AB}, \ref{tab:comp_exp_phon_ABprime} - Phonon frequencies at the $\Gamma$ point for the AB and AB' structures calculated with a selection of different methods.
    \item \ref{IR_theor_2_AB}, \ref{IR_theor_ratio_AB}, \ref{IR_theor_2_AB'}, \ref{IR_theor_ratio_ABprime} - Vibrational spectra for AB and AB' structures calculated with a selection of different methods.
    \item \ref{Rho_emp} - Sections of the valence charge density functions $\rho_{(PAW-PBE)}(\textbf{r})$ of fictitious systems obtained as explained in Sections III C and V E of the Main Text, used to calculate the data reported in Table \ref{epsilon_emp}.
    \item \ref{phonon_NA_emp} - Phonon energy dispersions calculated with the semi-empirical non-analytical corrections of Table \ref{epsilon_emp}.
    \item \ref{phonon_LDA} - Ultrasoft PP, LDA energy dispersions of phonons (NA: numerical implementation US-LDA DFPT).
    \item \ref{phonon_NC-PBE-TS} - NC PP, PBE, Tkatchenko-Scheffler corrected energy dispersions of phonons (NA: numerical implementation PAW-PBE [Set $a$] DFPT).
    \item \ref{Phonon_SCAN} - NC PP, SCAN energy dispersions of phonons (NA: numerical implementation NC-SCAN DFPT).
    \item \ref{phonon_PAW-PBE} - PAW PP, PBE, D2 vdW energy dispersions of phonons (NA: numerical implementation PAW-PBE [Set $b$] DFPT).
    \item \ref{AprimeB_661_0.03_0.15_NOvdW}, \ref{cell_par}, \ref{tab:comp_exp_phon_AB}, \ref{tab:comp_exp_phon_ABprime} - Tests for the effects of the formal vdW corrections.
    \item \ref{AA_661_0.10_0.11}, \ref{AprimeB_661_0.14_0.15}, \ref{AprimeB_661_0.03_0.15_NOvdW} - Non-standard explorations of the potential energy hypersurfaces.
    \item \ref{Conv_FD}, \ref{Conv_supercell_planar}, \ref{Conv_supercell_z} - Convergence studies.
\end{itemize}

\end{flushleft}
\centering
  \includegraphics[height=1cm]{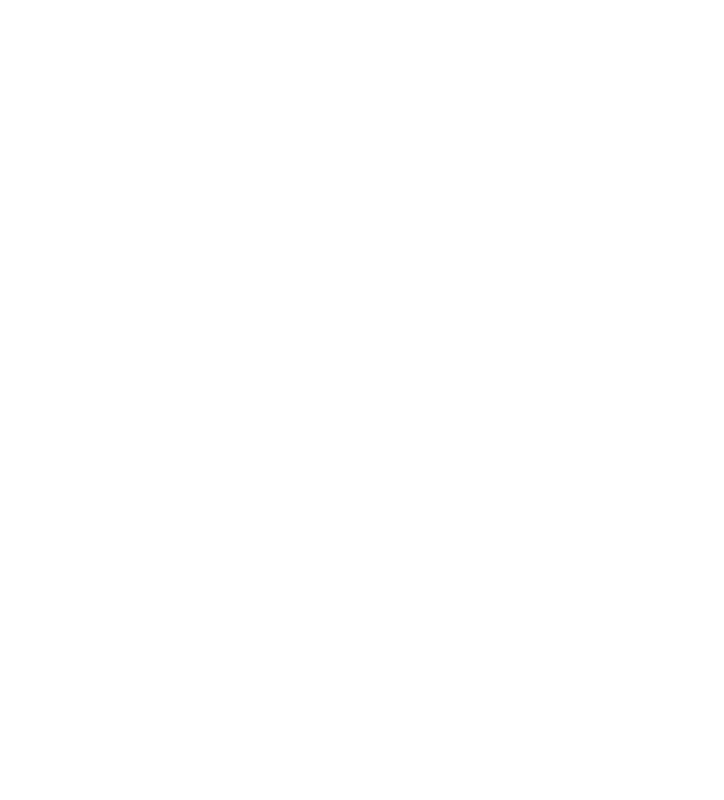}
\end{figure*}

\normalsize

\newcommand{\beginsupplement}{%
        \renewcommand{\thetable}{TS\Roman{table}}%
        \renewcommand{\thefigure}{S\arabic{figure}}%
           \setcounter{section}{0}
        \renewcommand{\thesection}{\Roman{section}}%
     }

\beginsupplement

\clearpage
\onecolumngrid

\section{Calculation parameters}

\begin{table*}[htb]
\centering
  \caption{Methods and calculation parameters used to obtain the results discussed in the text. Some of the employed settings, reported here, are motivated only by explorative purposes.}
  \label{tab:calc_param}
  \tiny{
  \setlength\tabcolsep{4.0pt} 
  \begin{tabular}{c|ccccc|cc|cccc}
    \hline Code &   \multicolumn{5}{c|}{General Calc. Setup}   &  \multicolumn{2}{c|}{SCF Calc. of $\rho_{x}(r)$} &  \multicolumn{4}{c}{FD IFC} \\
    & PP &xc  & \begin{tabular}[c]{@{}c@{}}Disp.\\ Correction\end{tabular} & \begin{tabular}[c]{@{}c@{}}$E_{cut.}$\\ (eV)\end{tabular}  &\begin{tabular}[c]{@{}c@{}}$E_{cut.}\rho$\\ (eV)\end{tabular}&\begin{tabular}[c]{@{}c@{}}\textbf{k}-points\\grid\end{tabular}&\begin{tabular}[c]{@{}c@{}}SCF\\ Conv. (eV)\end{tabular}&Supercell& \begin{tabular}[c]{@{}c@{}}\textbf{k}-points\\Grid\end{tabular} &Displ. (Å) &  \begin{tabular}[c]{@{}c@{}}SCF\\ Conv. (eV)\end{tabular}\\
    \hline
    \multirow{4}{*}{{\rotatebox[origin=c]{90}{QE}}} & NC  & PBE  & TS & 2041 & 8163 &  20$\times$20$\times$20 &  $1.36\cdot10^{-12}$  & \begin{tabular}[c]{@{}c@{}}AA': 6$\times$6$\times$2  \\ AB': 6$\times$6$\times$2 \\ Others: 5$\times$5$\times$2 \end{tabular}     & 2$\times$2$\times$2 & 0.01 & $1.36\cdot10^{-7}$ \\  \cline{2-12}
    & NC  & SCAN & -  & 2041 & 10885 & 20$\times$20$\times$20 &   $1.36\cdot10^{-12}$  &   6$\times$6$\times$2 & 2$\times$2$\times$2 & 0.01 & $1.36\cdot10^{-7}$\\   \cline{2-12}
    & US  & LDA  & -  & 544 &  10885 & 20$\times$20$\times$20 &   $1.36\cdot10^{-12}$    & \begin{tabular}[c]{@{}c@{}}AA': 6$\times$6$\times$2 \\A'B: 6$\times$6$\times$1\\ Others: 5$\times$5$\times$2\end{tabular}  & 2$\times$2$\times$2 & \begin{tabular}[c]{@{}c@{}} AB: 0.03\\ AB': 0.015\\Others: 0.01\end{tabular}  & $1.36\cdot10^{-7}$ \\    \cline{2-12}
    & \multirow{2}{*}{PAW} & \multirow{2}{*}{PBE}  &  \multirow{2}{*}{ D2 $^{Set \ a}$ }  & \multirow{2}{*}{544 }& \multirow{2}{*}{10885 } & \multirow{2}{*}{20$\times$20$\times$20} & \multirow{2}{*}{ $1.36\cdot10^{-12}$ } &  
    \begin{tabular}[c]{@{}c@{}}  AB': 6$\times$6$\times$1 \\ Others: 6$\times$6$\times$2\end{tabular}  &  2$\times$2$\times$2  
&\begin{tabular}[c]{@{}c@{}}AB': 0.15 \\ Others: 0.01\end{tabular}  &  $1.36\cdot10^{-7}$      
    \\ &&&&&&&& \multicolumn{4}{c}{AB': No disp. correction}\\
    \hline
    \multirow{2}{*}{\rotatebox[origin=c]{90}{VASP}} & PAW & PBE & \multicolumn{1}{c}{\begin{tabular}[c]{@{}c@{}}D2 $^{Set \ b}$ \\ D3BJ\\ TS\end{tabular}}  &   500  &  - & 12$\times$12$\times$4     & $1\cdot10^{-8}$   & 4$\times$4$\times$2 & 3$\times$3$\times$2 & 0.01 & $1\cdot10^{-8}$ \\\cline{2-12}
    & PAW & SCAN & \multicolumn{1}{c}{\begin{tabular}[c]{@{}c@{}}SCAN\\ +rVV10\end{tabular}} &   500  & - &  12$\times$12$\times$4  & $1\cdot10^{-8}$  & 4$\times$4$\times$2 & 3$\times$3$\times$2 & 0.01 & $1\cdot10^{-8}$\\
    \hline
  \end{tabular}
  \begin{flushleft}
    NC - Norm-conserving Troullier–Martins FHI PP \cite{fuchs1999ab, hamann1979norm,hamann2013optimized,troullier1991efficient,hamann2017erratum}, 
    US - Vanderbilt Ultrasoft PP \cite{vanderbilt1990soft}, 
    TS - Tkatchenko-Scheffler dispersion correction \cite{Tkatchenko2009}, 
    D2 - Grimme-D2 dispersion correction \cite{Grimme2006}, 
    PAW - Projector Augmented-Wave method \cite{Blochl_PhysRevB_50_1994,Kresse_PhysRevB_59_1999}, 
    SCAN - Strongly Constrained and Appropriately Normed semilocal density functional \cite{sun2016accurate,VASP_SCAN}, 
    SCAN+rVV10 - SCAN + revised Vydrov–van Voorhis nonlocal correlation functional \cite{VASP_SCANrVV10}, 
    D3BJ - Grimme-D3 dispersion correction with Becke-Johnson damping function \cite{Grimme2011}, 
    Set $a$ and $b$ - Internal references used in the text to indicate the specific settings reported in Rows 4 and 5 of this table.\\
    \end{flushleft}
    }
\end{table*}

\clearpage
\twocolumngrid

\section{Description of the potential energy surfaces}

\begin{figure*}[htb]
\textbf{AB' structure}\\
Potential energy surfaces\\
\centering
  \includegraphics[height=15.5cm]{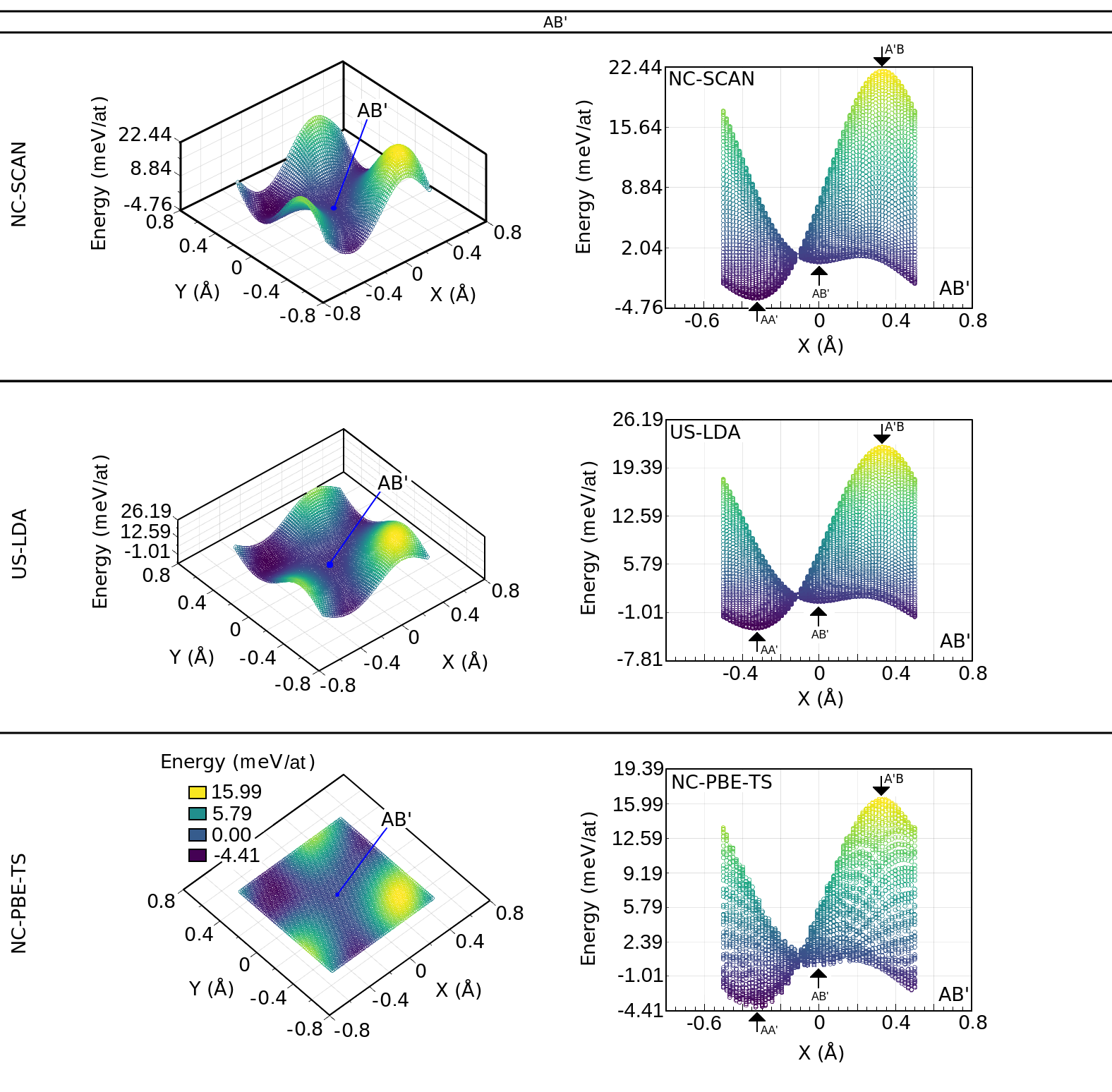}
  \caption{Potential energy surfaces (PESs) cuts calculated, with different theoretical implementations, upon rigid sliding movements of contiguous h-BN planes, keeping fixed the optimized cell structural parameters obtained for the \textbf{AB' symmetrical settlement} with each different method (reported in Table \ref{cell_par}). For Computational details see the relevant Section \ref{subsec:CompDetails} of the Main Text and Table \ref{tab:calc_param} of the Electronic Supplementary Information.}
  \label{ABprime_PES}
\end{figure*}

\small{
The potential energy surfaces (PESs) cuts reported in this work (Figure \ref{fig:PESstable} in the Main Text and Figure \ref{ABprime_PES} in the Electronic Supplementary Information [E.S.I.]) are calculated employing a selection of different methods and generated by parallel sliding of the contiguous h-BN planes with respect to each other, keeping the optimized lattice parameters and displacing only the atomic positions. The reported results span intervals from -0.5 Å to 0.5 Å in the $x$ and $y$ directions with respect to the symmetrical structures. 
Note that we used optimized structural parameters for each different plot (see E.S.I. [Table \ref{cell_par}]) and the absolute energies of the symmetrical points are, therefore, inconsistent among the plots themselves (the reason being that, e.g., the AA' structure obtained shifting from the fully optimized AB' comes in a different periodicity with respect to a fully relaxed AA').

The shape of the PES confirms the stability of the AA’ and AB symmetrical structures by being true minima, which is backed up by the comparison of the total energies.
The AB configuration appears to be the most stable one with an energy difference to the next most stable stacking between 0.22 meV/atom (NC-PBE-TS) to 2.04 meV/atom (PAW-PBE-TS). The PAW-PBE-TS predicted the AB' structure to be the most stable one while AB being second with an energy difference in the minimum of 0.76 meV/atom.
The total energy of the AB' configuration ranges from 1.81 meV/atom (NC-PBE-TS) to 5.37 meV/atom (QE-PAW-PBE-D2) above the AB minimum. The values calculated with QE-PAW-PBE-D2, and to a lesser extent NC-SCAN, significantly exceed the energy differences obtained from other methods.
The local minimum of AB' is located in a prominently flat valley (or a soft groove in NC-SCAN and US-LDA). Although the AA' configuration is  located at (-0.33,0.33) \AA\ from the AB' local minima and is easily reachable upon simple sliding, the AB' stacking configuration seems to be metastable, leading to possible dynamical stability related interpretations of the experimental data \cite{xu2019optomechanical,henck2017stacking}. This is also supported by the cozy shape of the PES in the vicinity of the AB' minimum 
particularly prominent in the NC-SCAN data and reported in more detail in the E.S.I. (Figure \ref{ABprime_PES}, uppermost panel).
It is not possible to get to the AB ($P3m1$) stacking by linearly sliding the planes from the AA' or AB' ($P6_{3}/mmc$) local minima. A 60$^{\circ}$ rotational movement is instead necessary, due to the different symmetry of the lattice space group.
The AA and A'B configurations are located on local maxima stationary points (Fig. \ref{fig:PESstable} of the Maint Text), implying instability of these stacking variants. The energy drop is strongly dependent on the theoretical methodology: the NC-PBE-TS method calculates significantly lower total energies for these unstable configurations, while the higher energy results are obtained with QE-PAW-PBE-D2.}
\\
\\
\\
\\
\\
\\
\\
\\
\\
\\
\\
\\
\\
\\
\\
\section{Description of the phononic modes}

\begin{figure*}
\textbf{Norm-conserving PP, PBE, Tkatchenko-Scheffler vdW energy dispersions of phonons}\\
NA: PAW-PBE (Set $a$) DFPT
\centering
  \includegraphics[height=8.5cm]{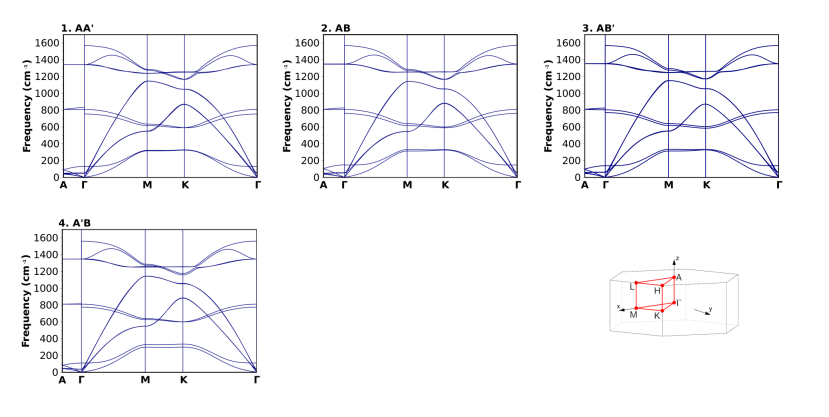}
  \caption{Comparison of the phonon energy dispersions along high symmetry directions in the first Brillouin zone of four differently stacked h-BN systems. The IFC matrices are calculated by Finite Displacements method with \textbf{DFT PBE functional, NC pseudopotential approximations and TS dispersion corrections}. The NA parts of the dynamical matrices are calculated adopting numerically the resulting PAW-PBE (Set $a$) DFPT effective charges and dielectric tensors. See further Computational details and theoretical information on Methods in the relevant Section \ref{subsec:CompDetails} of the Main Text and Table \ref{tab:calc_param} of the Electronic Supplementary Information.}
  \label{phonon_NC-PBE-TS}
\end{figure*}

\begin{figure*}
\textbf{PAW PP, PBE, D2 vdW energy dispersions of phonons} \\
\centering
  \includegraphics[height=3.5cm]{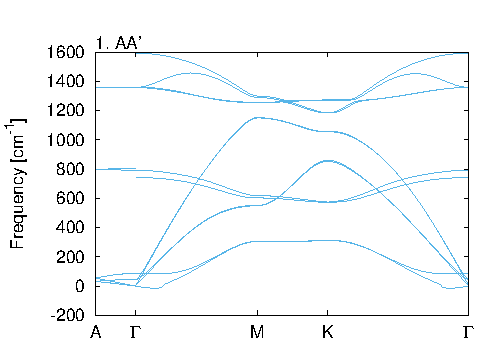}
  \includegraphics[height=3.5cm]{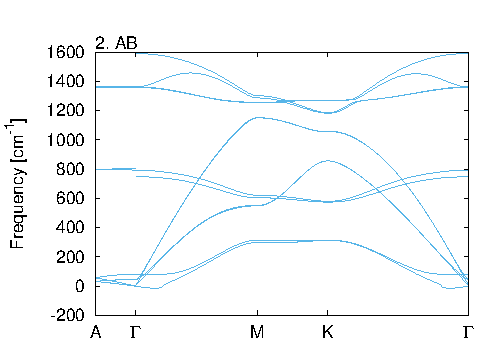}
  \includegraphics[height=3.5cm]{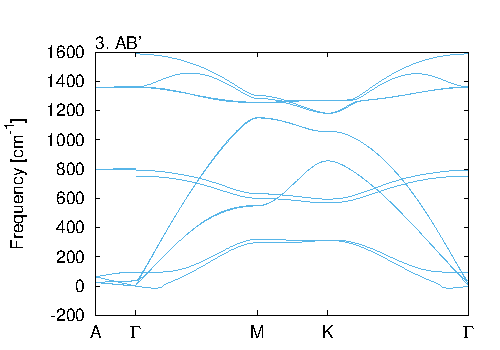}
  
  \includegraphics[height=3.5cm]{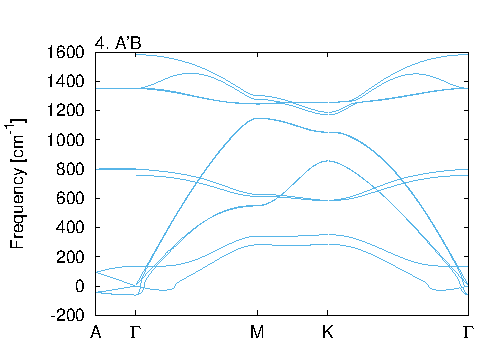}
  \includegraphics[height=3.5cm]{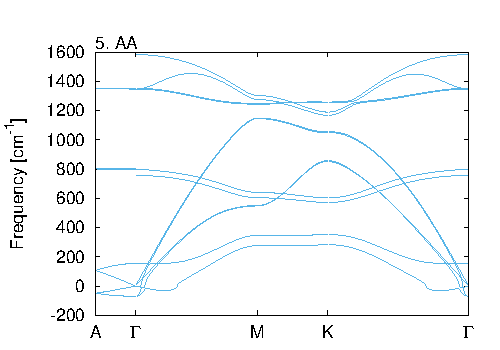}
  \includegraphics[height=3.5cm]{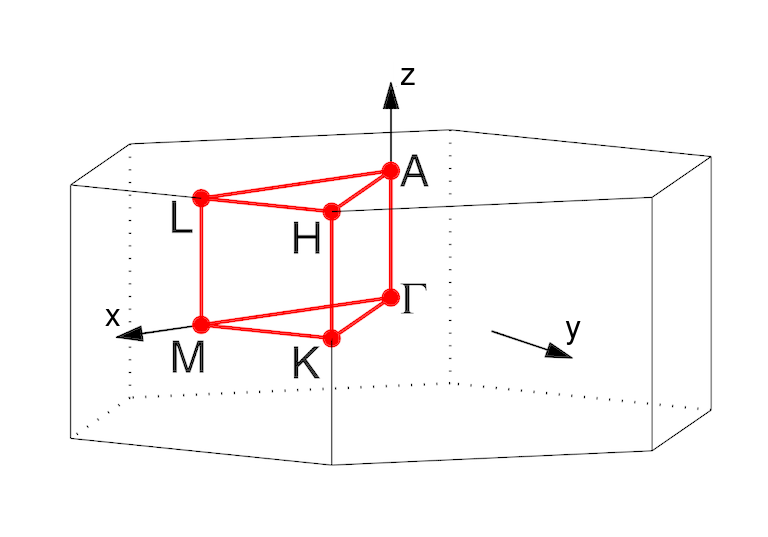}
  \caption{Comparison of the phonon energy dispersions along high symmetry directions in the first Brillouin zone for the five considered differently stacked h-BN systems. The IFC matrices are calculated by Finite Displacements method with \textbf{FD of 0.01 {\AA}, DFT PBE functionals, PAW pseudopotential approximations, Grimme-D2 dispersion corrections (VASP code. Set $b$ Table \ref{tab:calc_param})}. Cell parameters obtained with the same implementation PAW-PBE-D2 (Set $b$, Table \ref{cell_par}). The NA parts of the dynamical matrices are calculated adopting numerically the resulting PAW-PBE (Set $b$) DFPT effective charges and dielectric tensors. See further Computational details and theoretical information on Methods in the relevant Section \ref{subsec:CompDetails} of the Main Text and Table \ref{tab:calc_param} of the Electronic Supplementary Information. Interesting hints come from the comparison, in view of the different computational implementations, of these schemes with the ones reported in Fig. \ref{fig:phonon} of the Main Text. Keep in consideration the different FD lengths used for the AB' structure (and the AA and A'B non-standard explorations reported here).}
  \label{phonon_PAW-PBE}
\end{figure*}

\begin{figure*}[ht]
  \textbf{Convergence study for the finite displacement value in FD SCF calculations}\\
AA' structure\\
\centering
  \includegraphics[height=4.5cm, width=6.5cm]{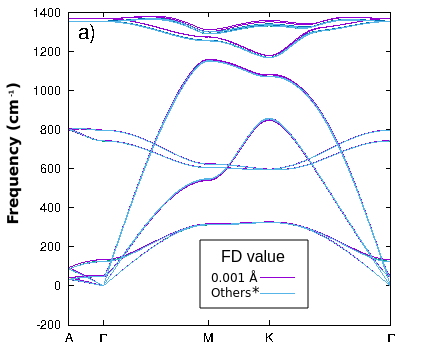}
  \includegraphics[height=5.0cm, width=6.2cm]{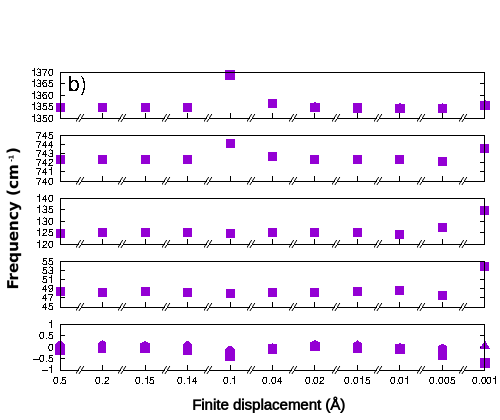}
  \caption{Evolution of the phonon frequencies (a) along high symmetry directions in the first Brillouin zone and (b) in the $\Gamma$ point, for the h-BN AA' structure (cell parameters obtained with PAW-PBE-D3BJ, reported in Table \ref{cell_par}, supercell size of $2\times2\times1$ unit cells), obtained with a set of different finite displacement values. Calculations performed with PAW-PBE and D3BJ van der Waals dispersion correction, see Computational details Section \ref{subsec:CompDetails} of the Main Text and Table \ref{tab:calc_param} of the Electronic Supplementary Information.\\ {*}Other tested values: 0.005 Å, 0.01 Å, 0.015 Å, 0.02 Å, 0.04 Å, 0.1 Å, 0.14 Å, 0.15 Å, 0.2 Å, 0.5 Å (the results are superimposable in Panel $a$, showed as light blue lines).}
  \label{Conv_FD}
\end{figure*}

\begin{figure*}[ht]
  \textbf{Convergence study for the $xy$ planar supercell size in FD SCF calculations}\\
AA' Structure\\
\centering
  \includegraphics[height=5.5cm, width=8.0cm]{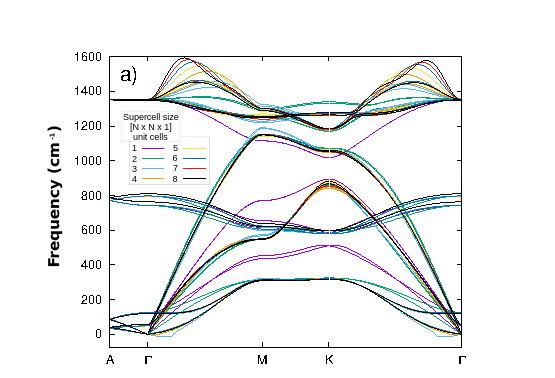}
  \includegraphics[height=4.8cm, width=7.0cm]{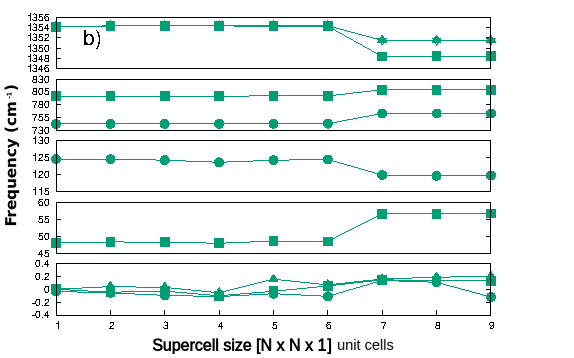}
  \caption{Phonon energy dispersions for the h-BN AA’ structure with varying supercell size in the in-plane \textit{xy} directions (a) along high symmetry directions in the first Brillouin zone and (b) in the $\Gamma$ point. Calculations performed with PAW-PBE and D3BJ van der Waals dispersion correction, FD value of 0.015 {\AA}, see further information in Computational details Section \ref{subsec:CompDetails} of the Main Text and Table \ref{tab:calc_param} of the Electronic Supplementary Information (cell parameters obtained with PAW-PBE-D3BJ, reported in Table \ref{cell_par}).}
  \label{Conv_supercell_planar}
\end{figure*}

\begin{figure*}[ht]
  \textbf{Convergence study for the $z$ planar supercell size in FD SCF calculations}\\
AA' Structure\\
\centering
  \includegraphics[height=6.8cm]{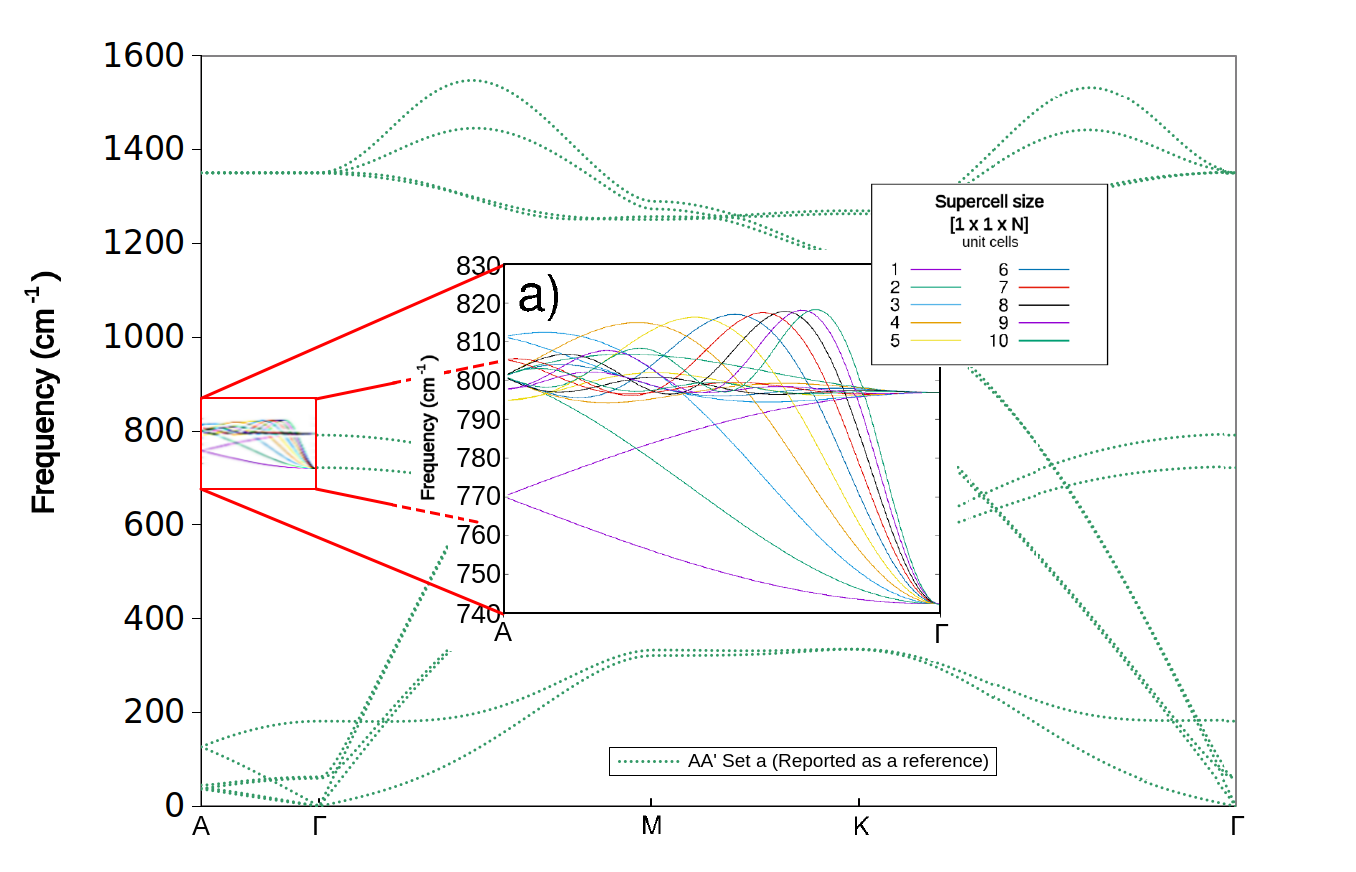}

  \includegraphics[height=3.5cm]{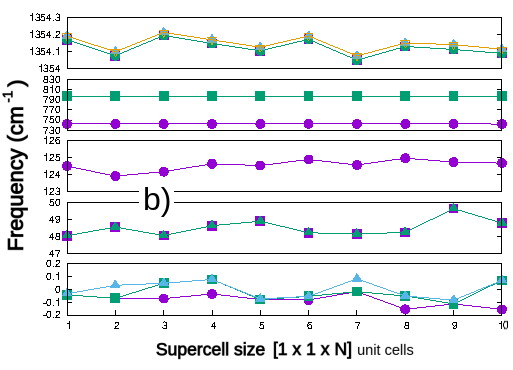}
  \includegraphics[height=3.5cm]{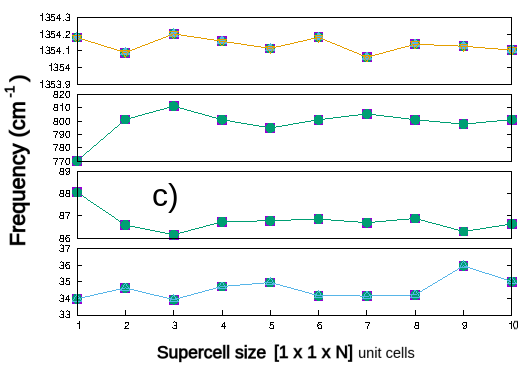}
  \caption{Phonon energy dispersions for the h-BN AA’ structure with varying  supercell size in the \textit{z} direction, perpendicular to the h-BN sheet, ($a$) along the $A-\Gamma$ high symmetry direction in the first Brillouin zone, ($b$) in the $\Gamma$ point and ($c$) in the $A$ point. Calculations performed with PAW-PBE and D3BJ van der Waals dispersion correction, FD value of 0.015 {\AA}, see further information in Computational details Section \ref{subsec:CompDetails} of the Main Text and Table \ref{tab:calc_param} of the Electronic Supplementary Information (cell parameters obtained with PAW-PBE-D3BJ, reported in Table \ref{cell_par}).}
  \label{Conv_supercell_z}
\end{figure*}

\small{
We calculated energy dispersion functions of the phononic modes for all the examined stacking variants with an extremely wide selection of methods. The results are shown in Fig. \ref{fig:phonon} of the Main Text and in the E.S.I.

Besides the three acoustic modes, the two plane sliding modes are easily recognizable below 100 cm$^{-1}$ ($E_{2g}$) at the $\Gamma$ point in all the dispersion diagrams, becoming imaginary in the unstable ones as indicated by the potential energy surfaces. 
A symmetrical "plane bumping" mode along the $z$ direction ($B_{1g}$) is distinguishable from 150 cm$^{-1}$ to 200 cm$^{-1}$. 
At about 800 cm$^{-1}$ at the $\Gamma$ point, two "umbrella movement" modes (symmetrical and asymmetrical) can be observed. The asymmetrical one ($A_{2u}$) is IR active and we denote it here as Peak 1. In Figure \ref{fig:abc_phonon} of the Main Text we report a graphical representation of the resulting displacements for the IR active modes at the $\Gamma$ point in the AA' system. The $A_{2u}$ mode is showed in Panel $c$. It consists by out-of-plane displacements of boron and nitrogen atoms in opposite directions (asymmetric bouncings) and interplanar van der Waals interactions are highly involved here. 
Finally, the higher energetic group of phonons is composed by four modes (higher than 1300 cm$^{-1}$ at the $\Gamma$ point) and arises from B-N stretching modes. The high energy results primarily from the deformation of in-plane covalent interactions. The $E_{2g}$ mode is symmetrical and Raman active while the $E_{1u}$ mode, composed by in-plane asymmetric stretching movements, is IR active and LO-TO split by the macroscopic electric field in two perpendicular branches. These two signals generally overlap each other in experimental spectra and we denote them, here, as a singular Peak 2. The displacement vectors relevant to it are represented in Panels $a$ and $b$ of Figure \ref{fig:abc_phonon} in the Main Text.

The stable and unstable structures can also be distinguished by the behaviour of the $B_{1g}$ and $A_{2u}$ modes (at $ \sim $ 800cm$^{-1}$ at the $\Gamma$ point) in the $A-\Gamma$ direction where the unstable structures exhibit a full degeneracy of the two modes, in contrast to the split seen in the stable stackings. The five stackings have otherwise an almost identical phonon structure. 
Concerning the IR and Raman active modes, the two most stable stackings (AA' and AB) exhibit identical phonon frequencies whereas the AB' presents slightly higher vibrational energies leading to different optical activity. 
We have calculated the phonon dispersion spectra with various dispersion correction methods (see Fig. \ref{fig:phonon} and Table \ref{tab:comp_exp_phon} in the Main Text and data reported in the E.S.I. [Tables \ref{tab:comp_exp_phon_AB}, \ref{tab:comp_exp_phon_ABprime}, Figures \ref{phonon_NC-PBE-TS}, \ref{phonon_PAW-PBE}, \ref{Conv_FD}, \ref{Conv_supercell_planar}, \ref{Conv_supercell_z}]). One would assume that due to the nature of the system in study different approaches to weak binding would have a significant impact on the phononic structure but this is not the case. The effect of the van der Walls correction can be seen only for the $B_{1g}$ mode arising from the planar shifts along the vertical axis $ \sim $ 120-200 cm$^{-1}$ as well as the $E_{2g}$ sliding of the planes along each other $ \sim $ 40-60 cm$^{-1}$. Both of these modes do not induce a change in the dipole moment and therefore will not be visible in the IR spectrum. When compared with experimental results, as shown in Tab. \ref{tab:comp_exp_phon} Main Text for the AA' variant and in the E.S.I. (Tables \ref{tab:comp_exp_phon_AB}, \ref{tab:comp_exp_phon_ABprime}) for the AB and AB', all the methods agree well on a qualitative level.}

\begin{table*}
\textbf{Lattice constants}
\caption{Lattice constants (in \AA) for the five analyzed conformations of h-BN, resulting from geometrical optimization with different theoretical approaches. Details about theoretical and computational methods are specified in the relevant sections: Methods Section (\ref{sec:Methods}) of the Main Text and Table \ref{tab:calc_param} of the Electronic Supplementary Information (E.S.I.).}
\label{cell_par}
\begin{tabular}{l|l|l|lllll}

\hline
& &                      & AA'          & AB   & AB'    & A'B     & AA \\
& & Method &$P6_{3}/mmc$ & $P3m1$ & $P6_{3}/mmc$ & $P6_{3}/mmc$ & $P\overline{6}m2$ \\
\hline

\multirow{8}{*}{a} & \multirow{4}{*}{\rotatebox[origin=c]{90}{VASP}} & PBE-D2 $^{1}$ ($b$)             & 2.509	&	2.508	&	2.508	&	2.508	&	2.508
\\
& & PBE-D3BJ   $^{1}$           & 2.507	&	2.507	&	2.506	&	2.506	&	2.506            \\
& & PBE-TS   $^{1}$             & 2.505	&	2.505	&	2.504	&	2.504	&	2.503            \\
& & SCAN+rVV10    $^{1}$        & 2.497	&	2.497	&	2.496	&	2.495	&	2.495            \\
\cline{2-8}
& \multirow{3}{*}{\rotatebox[origin=c]{90}{QE}} & PBE-D2    $^{1}$ ($a$)            & 2.513        & 2.513 & 2.510   & 2.510     & 2.510           \\
& & LDA    $^{2}$              & 2.490        & 2.490   & 2.490    & 2.490  & 2.490            \\
& & PBE-TS     $^{3}$          & 2.504        & 2.501   & 2.501    & 2.501  & 2.501            \\
& & SCAN    $^{3}$  & 2.506  &  2.506  & 2.504  &  2.504 &  2.504  \\
\hline
\hline
\multirow{8}{*}{c} & \multirow{4}{*}{\rotatebox[origin=c]{90}{VASP}} & PBE-D2 $^{1}$ ($b$)    & 6.159	&	6.096	&	6.214	&	6.608	&	6.699 \\
& & PBE-D3BJ  $^{1}$ & 6.567	&	6.547	&	6.606	&	6.927	&	6.968 \\
& & PBE-TS  $^{1}$   & 6.642	&	6.625	&	6.533	&	6.864	&	6.843 \\
& & SCAN+rVV10 $^{1}$ & 6.379	&	6.364	&	6.43	&	6.747	&	6.796 \\
\cline{2-8}
&  \multirow{4}{*}{\rotatebox[origin=c]{90}{QE}} & PBE-D2    $^{1}$ ($a$)      & 6.178 & 6.124 & 6.233  & 6.667 & 6.762  \\
& & LDA   $^{2}$      &6.490  & 6.450 & 6.531  & 7.074 & 7.155 \\
& & PBE-TS  $^{3}$    &6.653 & 6.613 & 6.708  & 7.277 & 7.386 \\
& & SCAN    $^{3}$  & 6.504 & 6.463  &  6.558 &  7.101  &  7.182 \\
\hline
\end{tabular}

$^{1}$ - Projector Augmented Wave PP \\
$^{2}$ - Ultrasoft PP\\
$^{3}$ - Norm-conserving PP

\end{table*}

\begin{table*}
\textbf{Atomic coordinates of the symmetrical structures}
\caption{Coordinates of the atomic centres in the five studied geometries. The coordinates are expressed in units of the hexagonal crystal lattice (reported in Tab. \ref{cell_par}).}
\label{tab:coordinates}

\begin{tabular}{l|l|lll}
 & & \textit{x} & \textit{y} & \textit{z} \\
 \hline
 \multirow{4}{*}{\rotatebox[origin=c]{90}{\textbf{AA'}}} & \textbf{B}   &  0.000000000  &  0.000000000  &  0.000000000\\
&\textbf{B}   & 0.333333333    &       0.666666667     &      0.500000000\\
&\textbf{N}   & 0.333333333      &     -0.333333333     &      0.000000000\\
&\textbf{N}  &  0.000000000    &     0.000000000     &      0.500000000\\

\hline

 \multirow{4}{*}{\rotatebox[origin=c]{90}{\textbf{AB}}}
&\textbf{B}  &   0.000000000    &       0.000000000      &     0.000000000\\
&\textbf{B}  &   0.666666667     &      -0.666666667      &     0.500000000\\
&\textbf{N}  &   0.333333333      &     -0.333333333      &     0.000000000\\
&\textbf{N}  &   0.000000000    &       0.000000000     &      0.500000000\\

\hline

 \multirow{4}{*}{\rotatebox[origin=c]{90}{\textbf{AB'}}}
&\textbf{B}   &  0.000000000     &    0.000000000  &       0.000000000\\
&\textbf{B}   &  0.000000000    &     0.000000000    &     0.500000000\\
&\textbf{N}   &  0.333333333     &    -0.333333333  &       0.000000000\\
&\textbf{N}   &  0.666666667    &     -0.666666667    &     0.500000000\\

\hline

 \multirow{4}{*}{\rotatebox[origin=c]{90}{\textbf{A'B}}}
&\textbf{B}  &  0.666666667      &   -0.666666667    &     0.500000000\\
&\textbf{B}  &  0.333333333  &      -0.333333333     &    0.000000000\\
&\textbf{N}  &   0.000000000    &     0.000000000  &       0.000000000\\
&\textbf{N}  &  0.000000000   &      0.000000000       &  0.500000000\\

\hline

 \multirow{4}{*}{\rotatebox[origin=c]{90}{\textbf{AA}}}
&\textbf{B}  &   0.000000000    &     0.000000000    &     0.000000000\\
&\textbf{B}  &  0.000000000     &    0.000000000    &     0.500000000\\
&\textbf{N} &   0.333333333      &   -0.333333333      &   0.000000000\\
&\textbf{N} &   0.333333333      &   0.666666667     &    0.500000000\\

\hline

\end{tabular}

\end{table*}



\begin{figure}
\textbf{AB structure}\\ Calculated infrared optical spectra 
\centering
  \includegraphics[height=12cm]{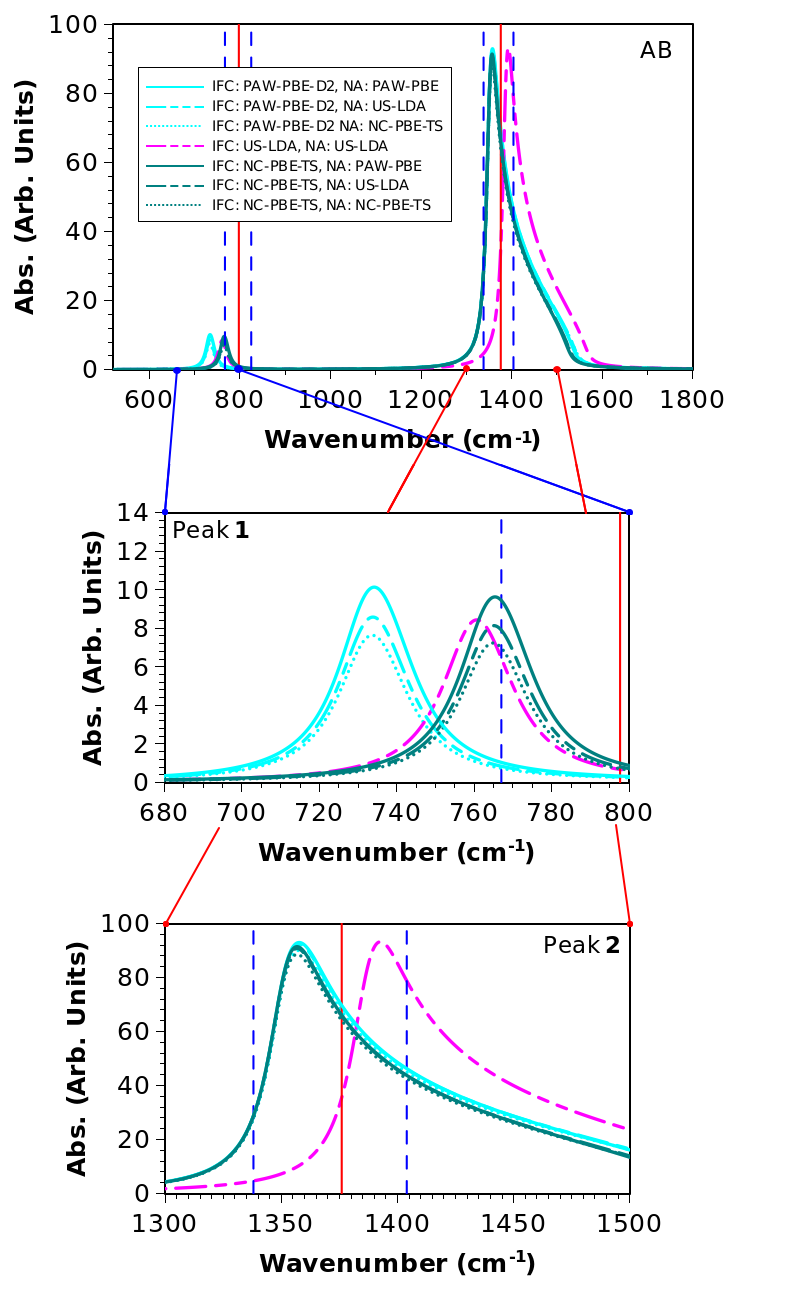}
  \caption{Vibrational spectra calculated for the h-BN \textbf{AB stacking configuration} with different theoretical approaches, as described in legend (for PAW-PBE we use Set $a$). In insets, details of Peak 1 and Peak 2 regions are showed. Red solid vertical lines indicate the average experimental peak values and blue vertical dotted lines indicate the range limits of measured high absorptions in the six experimental references \cite{ccamurlu2016modification,hidalgo2013high,mukheem2019boron,chen2017thermal,wang2003synthesis,andujar1998plasma}. Computational details and Methods are thoroughly examined in the relevant Section \ref{subsec:CompDetails} of the Main Text and Table \ref{tab:calc_param} of the Electronic Supplementary Information.}
  \label{IR_theor_2_AB}
\end{figure}
\begin{figure}[!ht]
\centering
  \includegraphics[height=5cm]{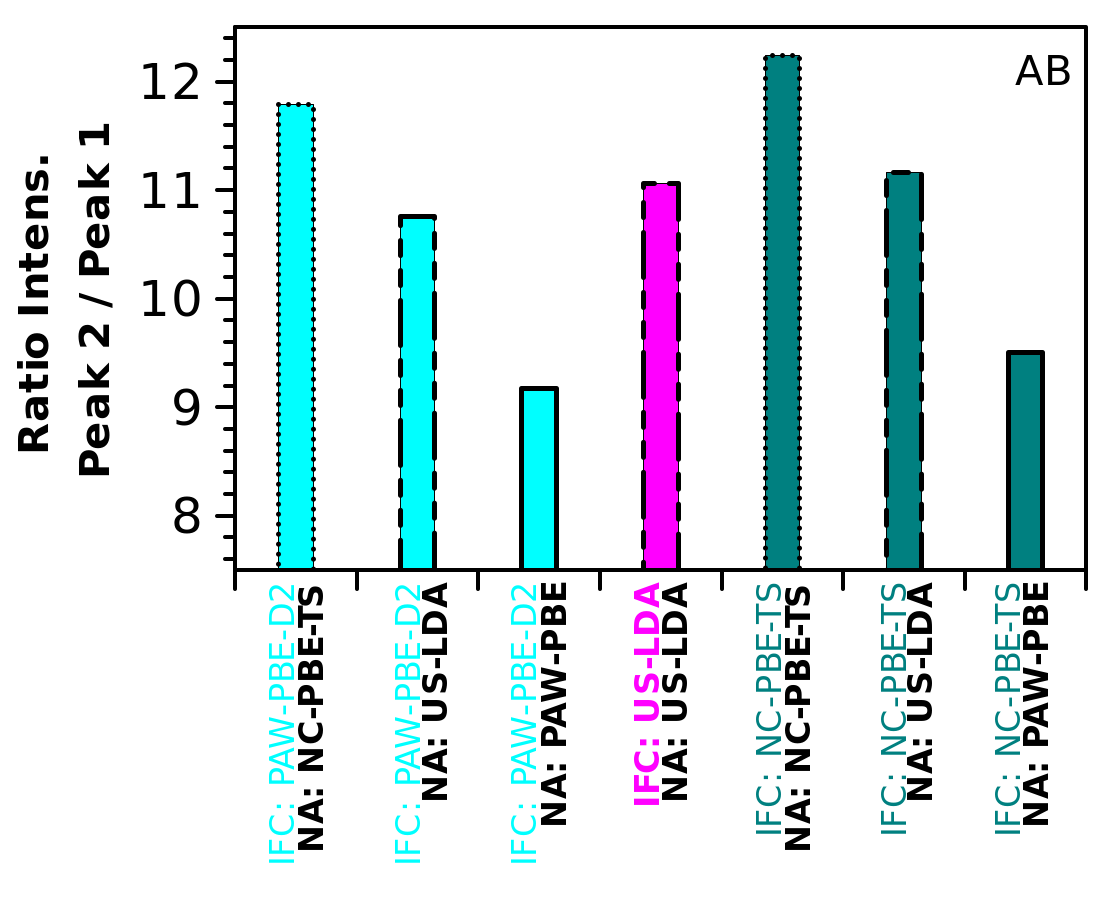}
  \caption{Ratios between the intensities of absorption of the two IR active peaks of vibrational spectra calculated for the simulated h-BN \textbf{AB stacking structure} with different theoretical implementations (for PAW-PBE we use Set $a$). Computational details and theoretical information on Methods are provided in the relevant Section \ref{subsec:CompDetails} of the Main Text and Table \ref{tab:calc_param} of the Electronic Supplementary Information.}
  \label{IR_theor_ratio_AB}
\end{figure}

\begin{figure}
\textbf{AB' structure}\\ Calculated infrared optical spectra 
\centering
  \includegraphics[height=12cm]{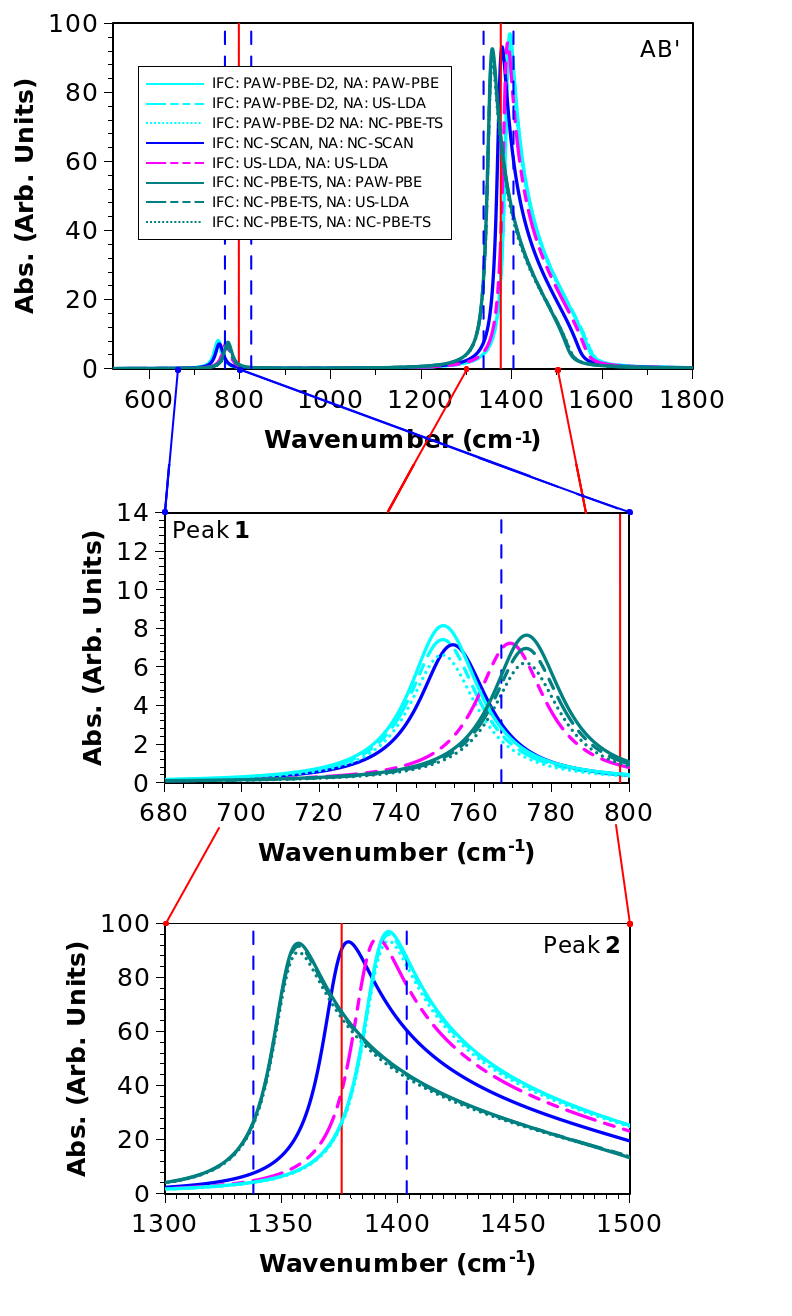}
  \caption{Vibrational spectra calculated for the h-BN \textbf{AB' stacking configuration} with different theoretical approaches, as described in legend (for PAW-PBE we use Set $a$). In insets, details of Peak 1 and Peak 2 regions are showed. Red solid vertical lines indicate the average experimental peak values and blue vertical dotted lines indicate the range of the experimental data for the two peak frequencies, as gathered from the six experimental references \cite{ccamurlu2016modification,hidalgo2013high,mukheem2019boron,chen2017thermal,wang2003synthesis,andujar1998plasma}. Computational details and theoretical information are provided in Section \ref{subsec:CompDetails} of the Main Text and Table \ref{tab:calc_param} of the Electronic Supplementary Information.}
  \label{IR_theor_2_AB'}
\end{figure}
\begin{figure}[!ht]
\centering
  \includegraphics[height=5cm]{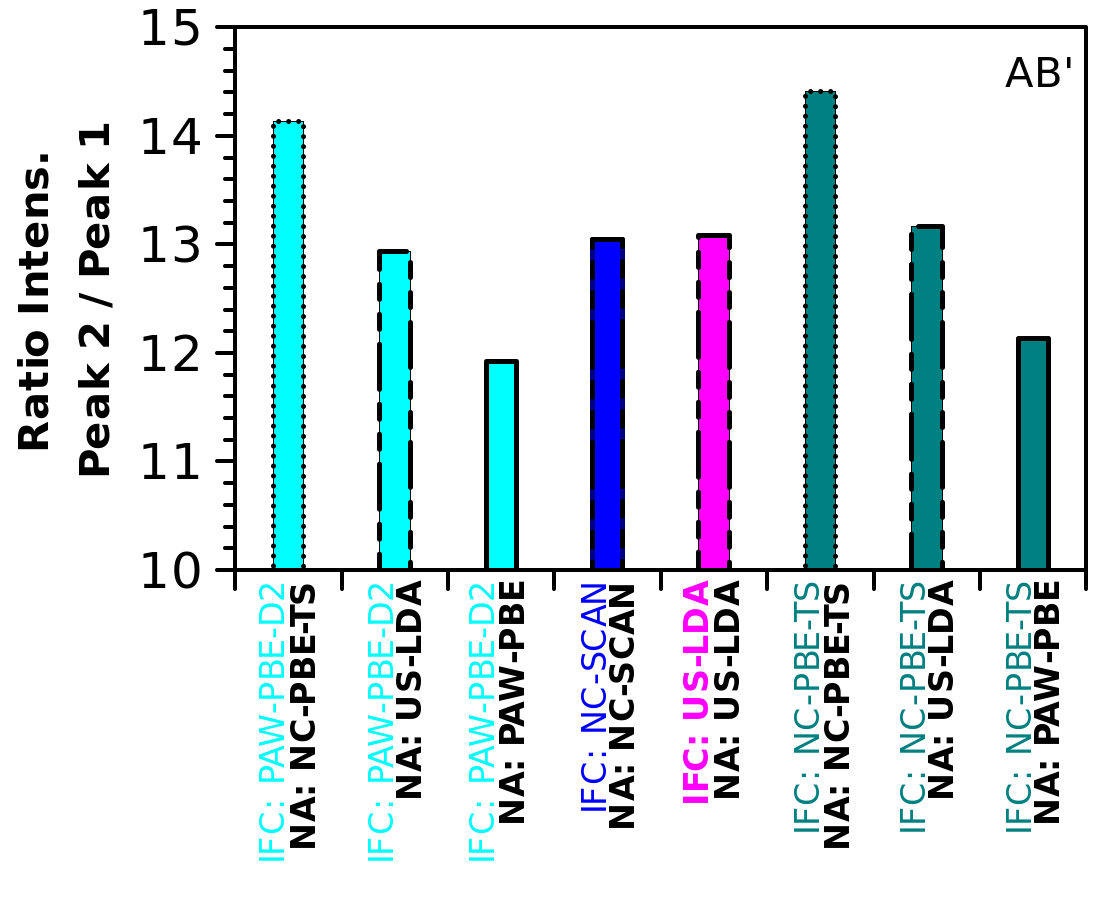}
  \caption{Ratios between the intensities of absorption of the two IR active peaks of vibrational spectra calculated for the simulated h-BN \textbf{AB' stacking structure} with different theoretical implementations (for PAW-PBE we use Set $a$). See the Computational details and Methods Section \ref{subsec:CompDetails} of the Main Text and Table \ref{tab:calc_param} of the Electronic Supplementary Information.}
  \label{IR_theor_ratio_ABprime}
\end{figure}

\begin{figure*}
\textbf{Charge density functions $\rho_{(PAW-PBE)}^{fict.}(\textbf{r})$ of the h-BN fictitious systems}, $\textbf{r}   \equiv  (R, \theta, \phi)$.\\ Used to obtain the semi-empirical Born $Q$ matrices and dielectric tensors reported in Table \ref{epsilon_emp}. \\ $\big[R < R_{HP}\big] \land  \big[{ \theta \leq \theta_{HP} }\big]  \implies   \rho_{(PAW-PBE)}^{fict.}(\textbf{r}) \approx  \rho_{(Q)}(\textbf{r}) $,\\ see Sections \ref{subsec:EmpDeriv} and \ref{subsec:TheorDescr} of the Main Text of this work.\\
\centering
  \includegraphics[height=13.5cm]{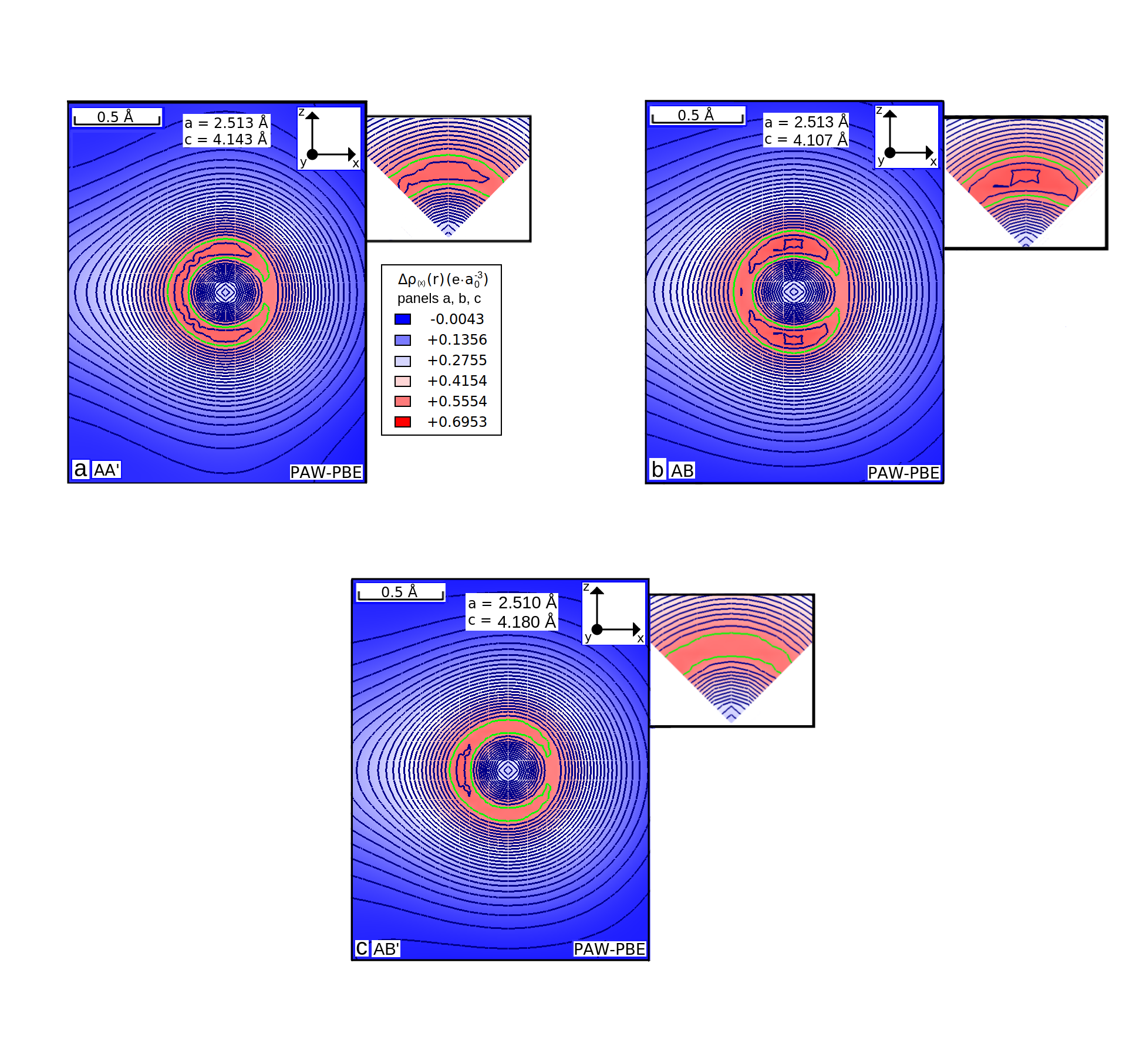}
   \caption{Nitrogen $xz$ sections of the valence charge density functions $\rho_{(PAW-PBE)}^{fict.}(\textbf{r})$ (Set $a$, recalculated with a $40\times 30 \times 20$ Monkhorst-Pack \textbf{k}-points grid in the first Brillouin zone of an equivalent orthorhombic unit cell, see Methods section \ref{subsec:CompDetails} in the Main Text and Table \ref{tab:calc_param} of the E.S.I.) calculated in h-BN fictitious systems obtained as explained in Sections \ref{subsec:TheorDescr} and \ref{subsec:EmpDeriv} of the Main Text, by applying the 32.94\% shortening percentage to the $c$ parameter in the $z$ direction from fully optimized (PAW-PBE-D2 [Set $a$]) h-BN structures (AA', AB and AB') and no further geometrical optimization (nuclei in relative positions). Nitrogen nuclei in the centres of the panels. The 0.5718 $e \cdot a_{0}^{-3}$ isolines of charge density are highlighted in green. In insets, details of polar cones sections. Close-up details are reported in Figure \ref{Rho_detail}.}
  \label{Rho_emp}
\end{figure*}

\begin{figure*}
\centering
  \includegraphics[height=7.0cm]{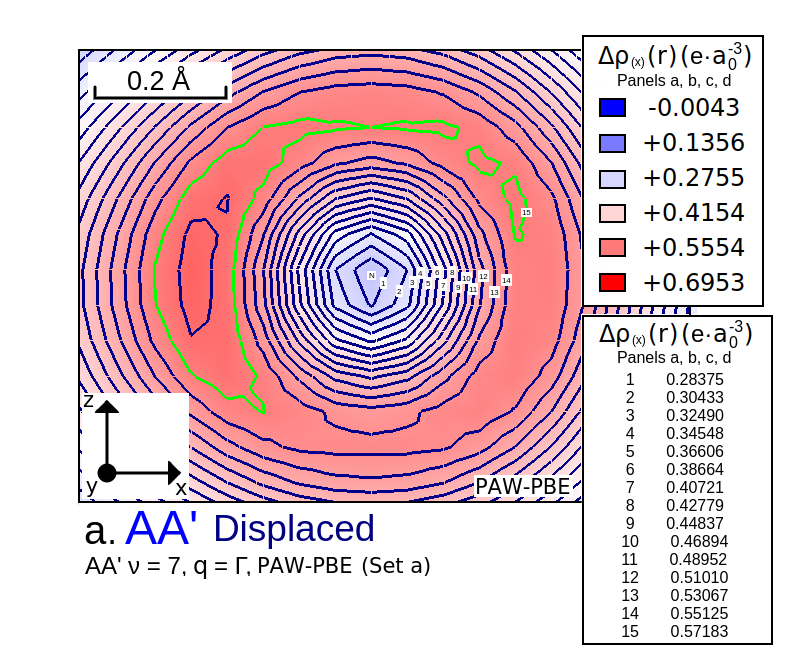}
  \includegraphics[height=6.9cm]{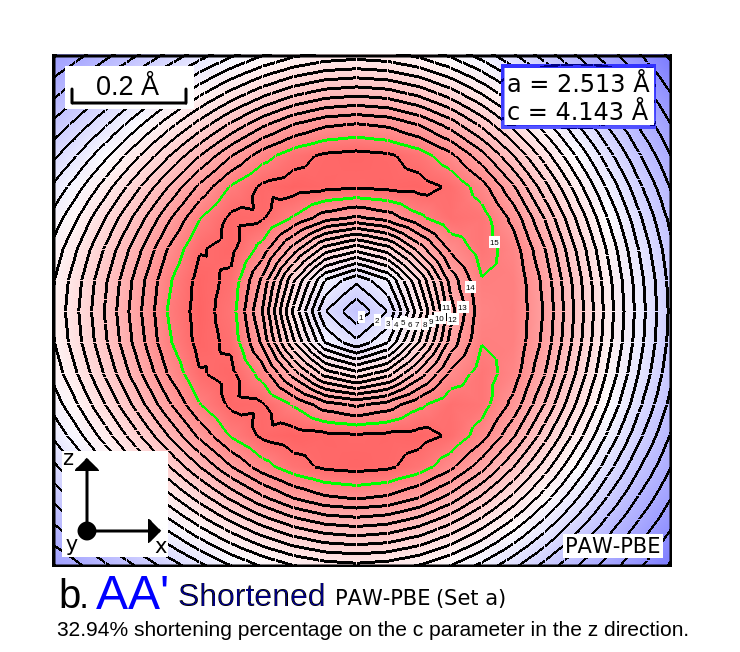}
  \includegraphics[height=7.0cm]{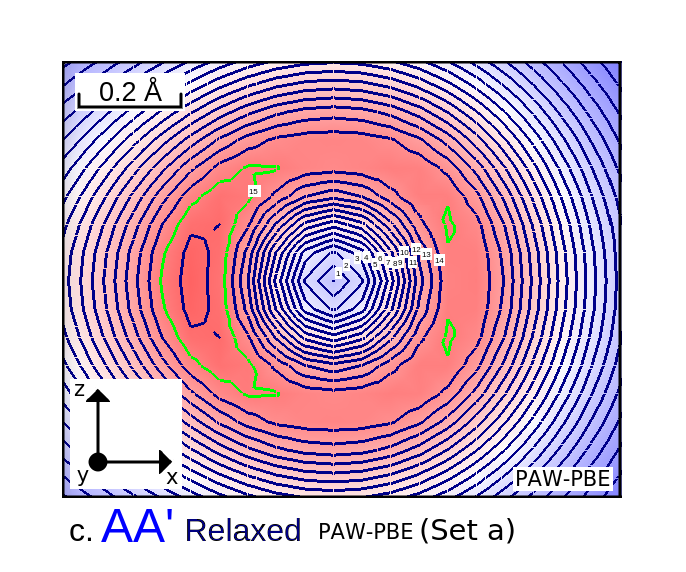}
  \includegraphics[height=7.0cm]{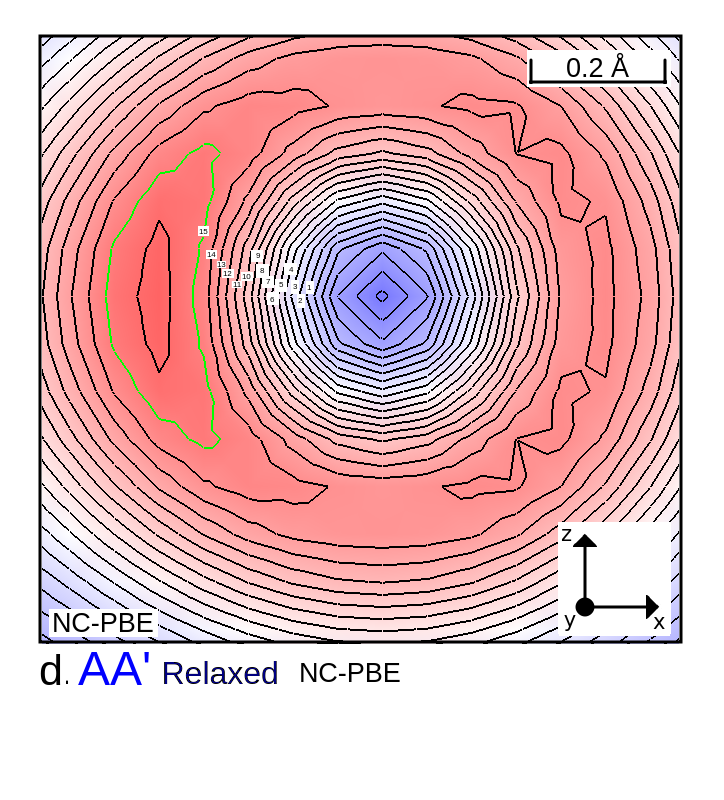}
   \caption{Nitrogen $xz$ sections, detailed maps for a set of real space valence charge density functions in selected h-BN AA' structures. Nitrogen nuclei are in the centres of the contour lines zones, approximately in the middle of the panels. The 0.5718 $e \cdot a_{0}^{-3}$ isolines of charge density are highlighted in green. The legends attached to Panel $a$ inform about the four panels.}
  \label{Rho_detail}
\end{figure*}

\begin{figure*}[htb]
\textbf{Phonon energy dispersions with semi-empirical non-analytical corrections}\\
Semi-empirical NA parameters reported in Table \ref{epsilon_emp}
\centering
  \includegraphics[height=4.3cm]{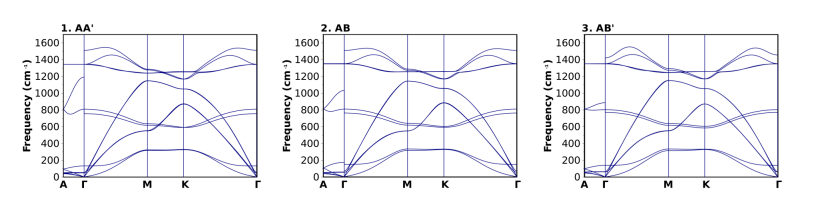}
  \caption{Comparison of the phonon energy dispersions along high symmetry directions in the first Brillouin zone for the three stable h-BN systems. The IFC matrices are calculated by Finite Displacements method with DFT PBE functional, NC pseudopotential approximations and TS dispersion corrections (see Computational details Section \ref{subsec:CompDetails} of the Main Text and Table \ref{tab:calc_param} of the Electronic Supplementary Information). \textbf{The NA parts (reported in Table \ref{epsilon_emp}) are semi-empirically obtained by DFPT PAW-PBE (Set $a$, see Computational details Section \ref{subsec:CompDetails} of the Main Text and Table \ref{tab:calc_param} of the Electronic Supplementary Information) on fictitious systems shortened on the $c$ parameter of 32.94\% with respect to the optimized PAW-PBE-D2 (Set $a$) geometry and no further geometrical optimization from PAW-PBE-D2 (Set $a$) optimal} (atomic centers in relative positions with respect to the cell parameters). See Sections \ref{subsec:EmpDeriv} and \ref{subsec:TheorDescr} of the Main Text of this work.}
  \label{phonon_NA_emp}
\end{figure*}

\begin{figure*}
\textbf{Ultrasoft PP, LDA energy dispersions of phonons}
\centering
  \includegraphics[height=8.5cm]{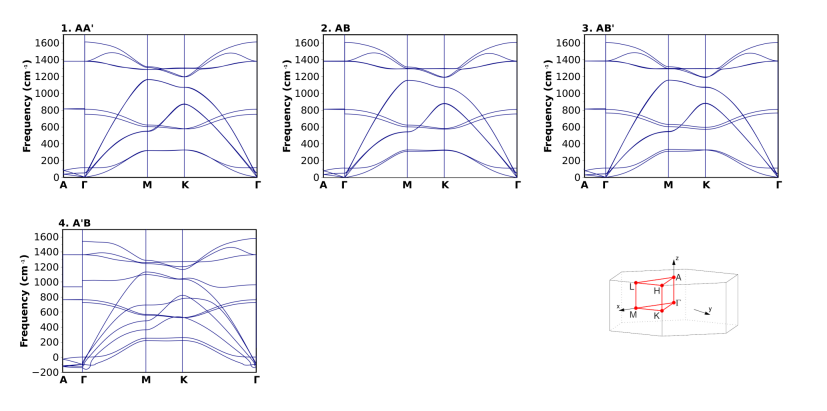}
  \caption{Comparison of the phonon energy dispersions along high symmetry directions in the first Brillouin zone for four differently stacked h-BN systems. The IFC matrices are calculated by Finite Displacements method with \textbf{DFT LDA functional, US pseudopotential approximations}. The NA parts of the dynamical matrices are calculated adopting numerically US-LDA DFPT effective charges and dielectric tensors. Computational details and theoretical information are provided in the relevant Section \ref{subsec:CompDetails} of the Main Text and Table \ref{tab:calc_param} of the Electronic Supplementary Information.}
  \label{phonon_LDA}
\end{figure*}

\begin{figure*}
\textbf{Norm-conserving PP, SCAN energy dispersions of phonons}\\
\centering
  \includegraphics[height=4.3cm]{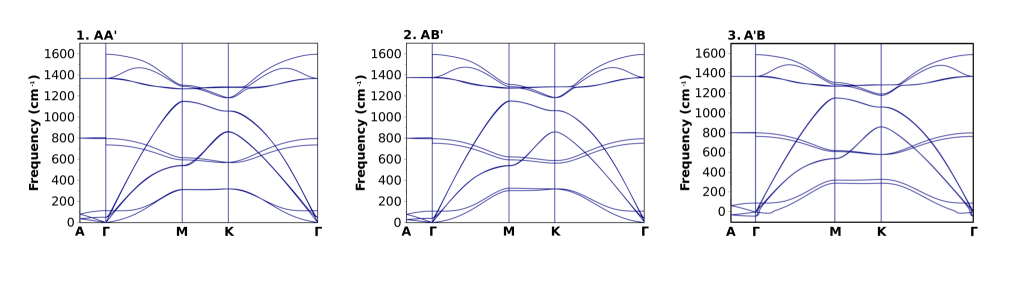}
  \caption{Comparison of the phonon energy dispersions along high symmetry directions in the first Brillouin zone of three differently stacked h-BN systems. The IFC matrices are calculated by Finite Displacements method with \textbf{DFT SCAN functional and NC pseudopotential approximations}. The NA parts of the dynamical matrices are calculated adopting numerically the results of NC-SCAN DFPT effective charges and dielectric tensors. See further Computational details and theoretical information on Methods in the relevant Section \ref{subsec:CompDetails} of the Main Text and Table \ref{tab:calc_param} of the Electronic Supplementary Information.}
  \label{Phonon_SCAN}
\end{figure*}

\begin{figure*}[htb]
\textbf{AA structure}\\
Non-standard explorations of the potential energy hypersurface 
\centering
  \includegraphics[height=5.2cm]{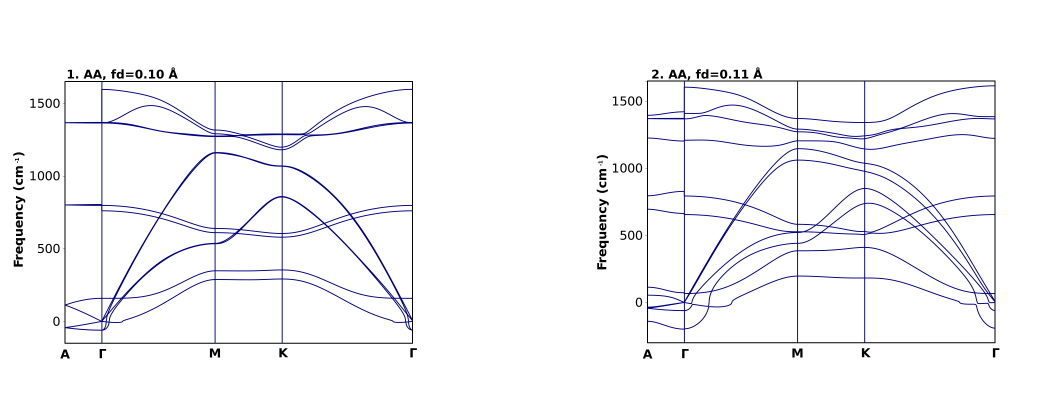}
  \caption{Non-standard explorations of the potential energy hypersurface of the h-BN \textbf{AA structure}. Phonon eigenvalues obtained from the diagonalization of a dynamical matrix in which the IFC results from long-range finite displacements. Supercell 6$\times$6$\times$1. \textbf{Finite displacements (1) 0.10 \AA, (2) 0.11 \AA}. Cell parameters obtained from PAW-PBE-D2 (Set $a$, Table \ref{cell_par}). \textbf{No van der Waals dispersion correction in the FD SCF calculations}. The NA part of the dynamical matrix is calculated adopting numerically PAW-PBE (Set $a$) DFPT effective charges and dielectric tensors. All the other parameters as reported in Table \ref{tab:calc_param}, Set $a$. Interesting hints come from the comparison of these unstable structures with the AB' scheme reported in Fig. \ref{fig:phonon} of the Main Text, obtained with the same theoretical implementation (Set $a$, Table \ref{tab:calc_param}) and long-range displacement of 0.15 {\AA}.}
  \label{AA_661_0.10_0.11}
\end{figure*}

\begin{figure*}[htb]
\textbf{A'B structure}\\
Non-standard explorations of the potential energy hypersurface 
\centering
  \includegraphics[height=5.5cm]{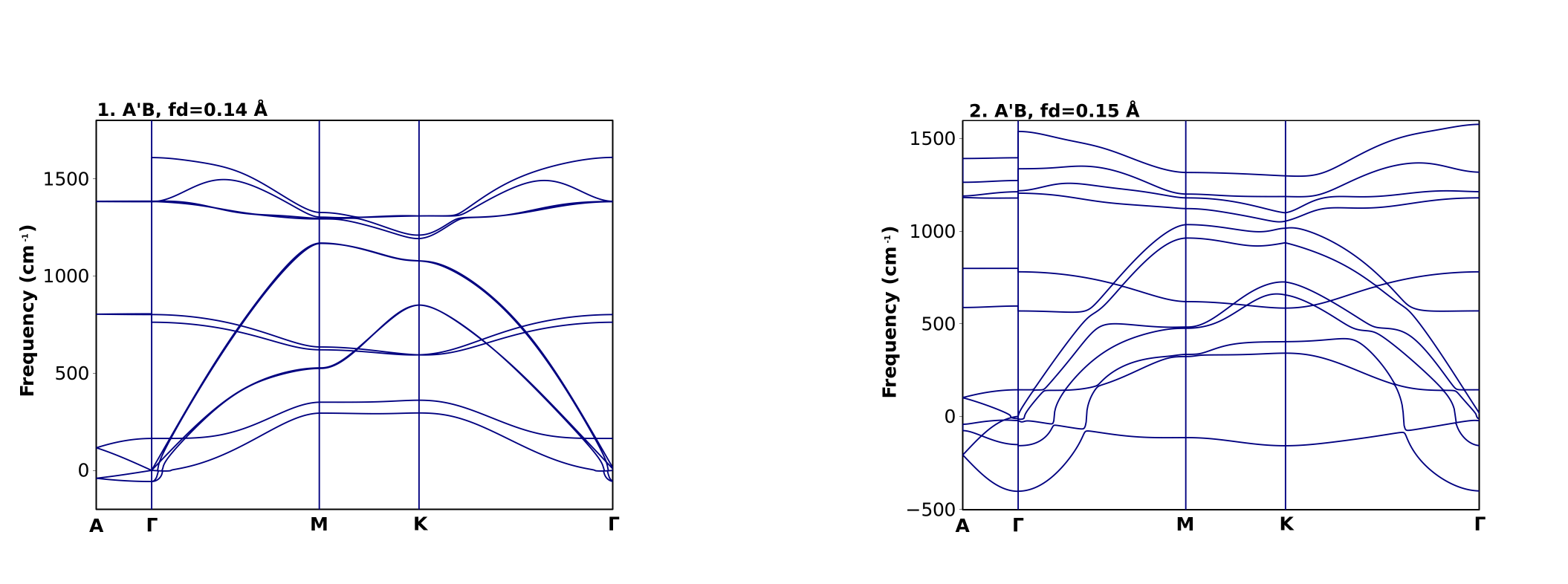}
  \caption{Non-standard explorations of the potential energy hypersurface of the h-BN \textbf{A'B structure}. Phonon eigenvalues obtained from the diagonalization of a dynamical matrix in which the IFC results from long-range finite displacements. Supercell 6$\times$6$\times$1. \textbf{Finite displacements (1) 0.14 \AA, (2) 0.15 \AA}. Cell parameters obtained from PAW-PBE-D2 (Set $a$, Table \ref{cell_par}). \textbf{No van der Waals dispersion correction in the FD SCF calculations}. The NA part of the dynamical matrix is calculated adopting numerically PAW-PBE (Set $a$) DFPT effective charges and dielectric tensors. All the other parameters as reported in Table \ref{tab:calc_param}, Set $a$. Interesting hints come from the comparison of these unstable structures with the AB' scheme reported in Fig. \ref{fig:phonon} of the Main Text, obtained with the same theoretical implementation (Set $a$, Table \ref{tab:calc_param}) and long-range displacement of 0.15 {\AA}.}
  \label{AprimeB_661_0.14_0.15}
\end{figure*}

\begin{figure*}[htb]
\textbf{A'B structure optimized without vdW corrections}\\
{Phonon energy dispersions}\\
with a non-standard exploration of the potential energy hypersurface

\centering
  \includegraphics[height=5.3cm]{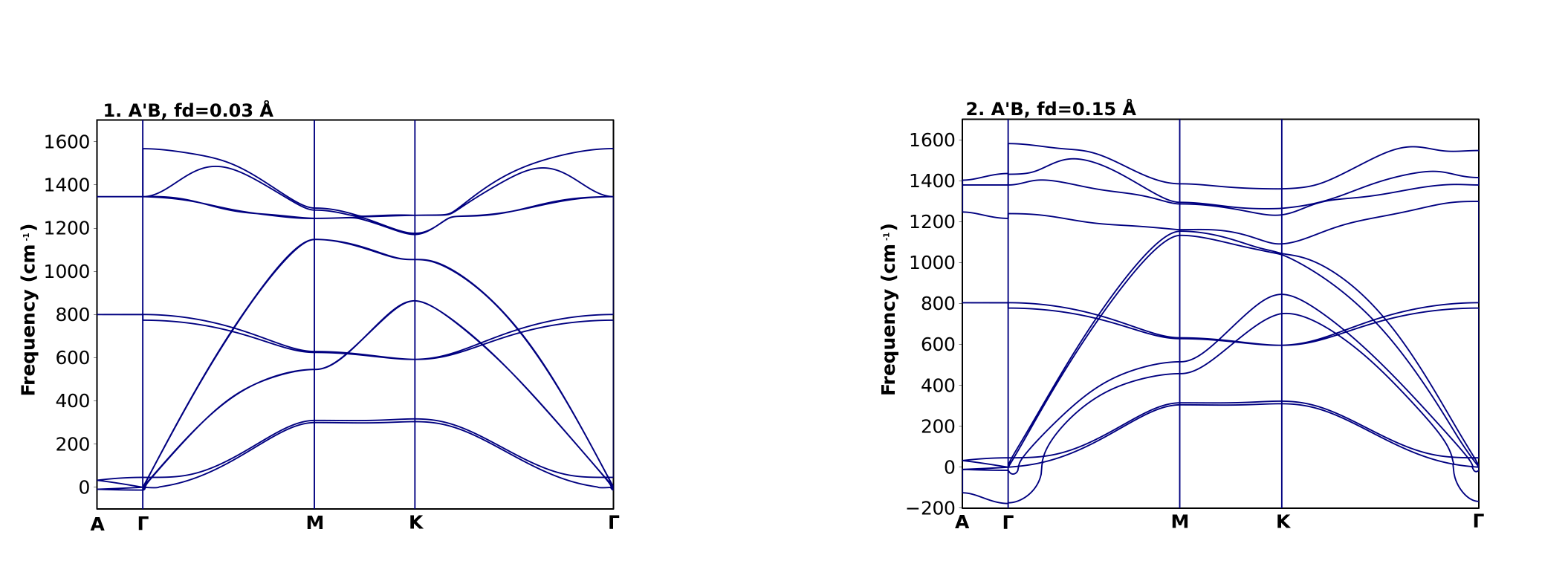}
  \caption{Phonon energy dispersions along high symmetry directions in the first Brillouin zone for the h-BN \textbf{A'B structure}. \textbf{Cell parameters obtained from PAW-PBE (Set $a$) calculations (without vdW dispersion corrections): a=2.51 \AA, c=8.20 \AA}. \textbf{No van der Waals dispersion correction in the FD SCF calculations}. Supercell 6$\times$6$\times$1. \textbf{Finite displacements: (1) 0.03 \AA, (2) 0.15 \AA}. The results reported in panel 2 represent a non-standard way to explore the potential energy hypersurface of the structure. In fact, the eigenvalues are obtained from the diagonalization of a dynamical matrix resulting from untypical long-range finite displacements (0.15 \AA). The NA part of the dynamical matrix is calculated adopting numerically PAW-PBE (Set $a$) DFPT effective charges and dielectric tensors. All the other parameters as reported in Table \ref{tab:calc_param} of the Electronic Supplementary Information, Set $a$.}
  \label{AprimeB_661_0.03_0.15_NOvdW}
\end{figure*}

\begin{table*}
\textbf{Semi-empirical Born $Q$ matrices and dielectric tensors}
    \centering
    \caption{Semi-empirically obtained non-analytical parts of the dynamical matrices for the three h-BN stable systems, calculated numerically with DFPT and PAW-PBE (Set $a$) on fictitious 32.94\% $c$ shortened systems. The shortening operation has been performed with respect to optimized geometry obtained with PAW-PBE-D2 (Set $a$, reported in Table \ref{cell_par}) and no further geometrical optimization (atomic centers in relative positions with respect to the cell parameters). See Sections \ref{subsec:EmpDeriv} and \ref{subsec:TheorDescr} of the Main Text of this work.}
    \label{epsilon_emp}

 \scriptsize{

\begin{tabular}{lll}

\begin{tabular}{llll}
\hline
 \textbf{AA'}  &  &   \\
\hline
 \multicolumn{4}{c}{ Dielectric tensor $\epsilon^{(\infty)}$  }  \\
\hline

 &  $x$   &    $y$     &  $z$\\

    $x$ &  5.410    &     0.000   &   0.000  \\
      $y$  &       0.000     &       5.410     &  0.000  \\
        $z$&     0.000   &       0.000    &       9.578  \\

\hline
   \multicolumn{4}{c}{Born $Q$ matrix elements $Q^{ \alpha   \beta}_{s}$ (e)}  \\
\hline
 &  $x$   &    $y$     &  $z$\\
     \multicolumn{3}{c}{ B (0,0,0) }  \\
          $x$ &    2.414     &       0.000       &    0.000  \\
          $y$  &   -0.022     &        2.426         &   0.000   \\
        $z$&    -0.008     &      0.000       &    4.284  \\
         \hline
    \multicolumn{3}{c}{  B (1/3,2/3,1/2) } \\
        $x$ &      2.414     &       0.000         &    0.000   \\
         $y$  &    -0.022      &       2.426        &   0.000   \\
        $z$&    -0.008      &     0.000         &   4.284  \\
         \hline
     \multicolumn{3}{c}{ N (1/3,-1/3,0)}   \\
        $x$ &     -2.412     &        0.000        &   0.000 \\
       $y$  &      -0.007      &      -2.409         &    0.000  \\
        $z$&    -0.002       &     0.000         &   -4.270  \\
       \hline
    \multicolumn{3}{c}{  N (0,0,1/2)   } \\
        $x$ &     -2.412     &      0.000      &   0.000  \\
         $y$  &    -0.007     &      -2.409        &    0.000   \\
         $z$&   -0.002     &      0.000        &   -4.270    \\

\hline
\end{tabular}

&

\begin{tabular}{llll}

\hline
 \textbf{AB}  &  &   \\
\hline
 \multicolumn{4}{c}{ Dielectric tensor $\epsilon^{(\infty)}$}  \\
\hline
 &  $x$   &    $y$     &  $z$\\

       $x$ &         7.010     &     0.000        &  0.000\\
          $y$  &   0.000    &      7.011         &0.000\\
       $z$&      0.000   &      0.000         &7.615\\

\hline
   \multicolumn{4}{c}{Born $Q$ matrix elements $Q^{ \alpha   \beta}_{s}$ (e)}  \\
\hline
 &  $x$   &    $y$     &  $z$\\
     \multicolumn{3}{c}{ B (0,0,0) }  \\
      $x$ &      2.351   &      0.000        &0.000\\
           $y$  &  -0.023    &      2.364       &0.000\\
       $z$&      -0.008    &     0.000         &3.552\\
         \hline
    \multicolumn{3}{c}{  B (2/3,-2/3,1/2) } \\
        $x$ &     2.992   &      0.000        &0.000\\
          $y$  & -0.025    &      3.007        &0.000\\
           $z$&   -0.009   &      0.000        &2.086\\
         \hline
     \multicolumn{3}{c}{ N (1/3,-1/3,0)}   \\
       $x$ &           -2.699    &     0.000 &     0.000\\
        $y$  &    -0.008      &   -2.695        &0.000\\
           $z$&   -0.003       &   0.000       &-2.656\\
       \hline
    \multicolumn{3}{c}{  N (0,0,1/2)   } \\
     $x$ &     -2.642    &    0.000          &0.000\\
    $y$  &     -0.004    &     -2.640        &0.000\\
       $z$&     -0.002       &   0.000       &-3.023\\

\hline
\end{tabular}

&

\begin{tabular}{llll}

\hline
 \textbf{AB'}  &  &   \\
\hline
 \multicolumn{4}{c}{ Dielectric tensor $\epsilon^{(\infty)}$ }  \\
\hline

 &  $x$   &    $y$     &  $z$\\

     $x$ &     67.994   &     -26.684       &0.000\\
     $y$  &    -26.684   &      37.182       &0.000\\
      $z$&        0.000    &     0.000       &9.046\\

\hline
   \multicolumn{4}{c}{Born $Q$ matrix elements $Q^{ \alpha   \beta}_{s}$ (e)}  \\
\hline
  &  $x$   &    $y$     &  $z$\\
     \multicolumn{3}{c}{ B (0,0,0) }  \\
     
           $x$ &        4.525     &    -0.402          &0.000\\
         $y$  &   -0.444    &      4.086         &0.000\\
      $z$&       -0.014     &     0.002          &1.958\\
         \hline
    \multicolumn{3}{c}{  B (0,0,1/2) } \\
    
         $x$ &         4.525     &    -0.402        & 0.000\\
          $y$  &  -0.444      &    4.086          &0.000\\
       $z$&     -0.014     &     0.002          &1.958\\
         \hline
     \multicolumn{3}{c}{ N (1/3,-1/3,0)}   \\
     
      $x$ &          -4.546     &     0.424      &   0.000\\
           $y$  & 0.420     &    -4.055         &0.000\\
       $z$&   -0.001     &    0.000        &-1.994\\
       \hline
    \multicolumn{3}{c}{  N (2/3,-2/3,1/2)   } \\
      
      $x$ &   -4.546   &       0.424        &0.000\\
          $y$     &   0.420   &      -4.055        &0.000\\
        $z$&  -0.001   &      0.000       &-1.994\\

\hline
\end{tabular}

\\
\end{tabular}
}

\end{table*}

\begin{table*}
\textbf{AB structure}\\
Phonon frequencies at the $\Gamma$ point
\centering
  \caption{Phonon frequencies at the $\Gamma$ point (in $cm^{-1}$) for the h-BN \textbf{AB stacking} as calculated by different theoretical approaches and software implementations and compared with experimental values taken from literature. The NA part of the dynamical matrix is calculated adopting numerically DFPT effective charges and dielectric tensors, where not differently specified, with the same method of the IFC. For Computational details see the relevant Methods Section (\ref{subsec:CompDetails}) of the Main Text and Table \ref{tab:calc_param} of the Electronic Supplementary Information.}
  \label{tab:comp_exp_phon_AB}
\tiny{
\begin{tabular}{c|ccccc|cccc|c}
\hline
 & \multicolumn{5}{c|}{VASP} & \multicolumn{4}{c|}{QE} & \multirow{3}{*}{Experiment} \\
Mode & PBE$^{1}$  & PBE-D2$^{1}$ ($b$)  & PBE-D3BJ$^{1}$ & PBE-TS$^{1}$   & \begin{tabular}[c]{@{}c@{}}SCAN\\ +rVV10\end{tabular}$^{1}$ & LDA$^{2}$ & PBE-D2$^{1}$ ($a$) & PBE-TS$^{3}$ &    \begin{tabular}[c]{@{}c@{}}PBE-TS$^{3}$\\NA: PAW-PBE$^{1}$\end{tabular}   \\
\hline
\multirow{2}{*}{E$_{2g}$**} & 45   & 45      & 45             & 43      & 103           & 53   & 71      & 53      & 53                   & \multirow{2}{*}{51 \cite{Nemanich1981}   }          \\
      & 45   & 45      & 45             & 43      & 107           & 53   & 71      & 57      & 57                   &                        \\
B$_{1g}$   & 125  & 81      & 118            & 150     & 107           & 113  & 177     & 149     & 149                  & -                      \\
A$_{2u}$*  & 751  & 748     & 750            & 755     & 766           & 758  & 731     & 763     & 763                  & 767-810 \cite{Geick1966,hidalgo2013high,ccamurlu2016modification,mukheem2019boron,chen2017thermal,wang2003synthesis,andujar1998plasma}   \\
B$_{1g}$   & 798  & 794     & 798            & 802     & 821           & 812  & 793     & 810     & 810                  & -                      \\
\multirow{2}{*}{E$_{2g}$**} & 1357 & 1359    & 1356           & 1343    & 1378          & 1386 & 1350    & 1350    & 1350                 & \multirow{2}{*}{1369-1376 \cite{Geick1966,Nemanich1981,Reich2005}}    \\
      & 1359 & 1362    & 1359           & 1346    & 1383          & 1390 & 1357    & 1354    & 1354                 &                        \\
E$_{1u}$*  & 1360 & 1362    & 1360           & 1346    & 1385          & 1389 & 1356    & 1353    & 1353                 & 1338-1404 \cite{Geick1966,hidalgo2013high,ccamurlu2016modification,mukheem2019boron,chen2017thermal,wang2003synthesis,andujar1998plasma} \\
E$_{1u}$*  & 1589 & 1591    & 1589           & 1576    & 1622          & 1613 & 1588    & 1576    & 1573                 & 1616 \cite{Geick1966}  \\         
\hline
\multicolumn{11}{l}{\begin{tabular}[c]{@{}l@{}}$^{1}$ - Projector Augmented-Wave PP, $^{2}$ - Ultrasoft PP, $^{3}$ - Norm-conserving PP\\ * - IR active modes \\ ** - Raman active modes\end{tabular}}
\end{tabular}
}

\end{table*}

\begin{table*}
\textbf{AB' structure}\\
Phonon frequencies at the $\Gamma$ point
\centering
  \caption{Phonon frequencies at the $\Gamma$ point (in $cm^{-1}$) for the h-BN \textbf{AB' stacking} as calculated by different theoretical approaches and software implementations and compared with experimental values taken from literature. The NA part of the dynamical matrix is calculated adopting numerically DFPT effective charges and dielectric tensors, where not differently specified, with the same method of the IFC. See the relevant Methods Section (\ref{subsec:CompDetails}) of the Main Text and Table \ref{tab:calc_param} of the E.S.I. for computational details.}
  \label{tab:comp_exp_phon_ABprime}
\tiny{
\begin{tabular}{l|rrrrr|rrrrr|c}

\hline
     & \multicolumn{5}{c|}{VASP}         & \multicolumn{5}{c|}{QE} & \multirow{2}{*}{Experiment} \\
Mode & PBE$^{1}$  & PBE-D2$^{1}$ ($b$)  & PBE-D3BJ$^{1}$ & PBE-TS$^{1}$   & \begin{tabular}[c]{@{}c@{}}SCAN\\ +rVV10\end{tabular}$^{1}$ & LDA$^{2}$ & PBE-D2$^{1}$ ($a$) & PBE-TS$^{3}$ &    \begin{tabular}[c]{@{}c@{}}PBE-TS$^{3}$\\NA: PAW-PBE$^{1}$\end{tabular} &  SCAN$^{3}$  \\
\hline
\multirow{2}{*}{E$_{2g}$**} & 36   & 35   & 36   & 45   & 121  & 40   & 51   & 60   & 60   &  40 & \multirow{2}{*}{51 \cite{Nemanich1981}}                                                                                                                   \\
           & 36   & 35   & 36   & 45   & 121  & 41   & 52   & 62   & 62   &   41                                                                                                                                       \\
B$_{1g}$   & 128  & 95   & 121  & 154  & 153  & 110  & 174  & 138  & 138  & 109   &  -                                                                                                                                        \\
A$_{2u}$*  & 756  & 752  & 756  & 761  & 774  & 767  & 750  & 771  & 771  &   752   &767-810 \cite{Geick1966,hidalgo2013high,ccamurlu2016modification,mukheem2019boron,chen2017thermal,wang2003synthesis,andujar1998plasma}   \\
B$_{1g}$   & 796  & 794  & 795  & 795  & 824  & 810  & 794  & 804  & 804  &  795   & -                                                                                                                                        \\
\multirow{2}{*}{E$_{2g}$**} & 1357 & 1360 & 1357 & 1359 & 1379 & 1386 & 1390 & 1352 & 1352 & 1373  & \multirow{2}{*}{1369-1376 \cite{Geick1966,Nemanich1981,Reich2005}   }                                                                                     \\
           & 1360 & 1362 & 1360 & 1362 & 1379 & 1389 & 1396 & 1354 & 1354 &   1376   &                                                                                                                                    \\
E$_{1u}$*  & 1360 & 1362 & 1360 & 1362 & 1380 & 1389 & 1394 & 1354 & 1354 &  1376     &1338-1404 \cite{Geick1966,hidalgo2013high,ccamurlu2016modification,mukheem2019boron,chen2017thermal,wang2003synthesis,andujar1998plasma} \\
E$_{1u}$*  & 1586 & 1588 & 1586 & 1588 & 1620 & 1608 & 1616 & 1575 & 1569 &   1594   &1616 \cite{Geick1966}   \\ 
\hline
\multicolumn{11}{l}{\begin{tabular}[c]{@{}l@{}}$^{1}$ - Projector Augmented-Wave PP, $^{2}$ - Ultrasoft PP, $^{3}$ - Norm-conserving PP\\ * - IR active modes \\ ** - Raman active modes\end{tabular}}
\end{tabular}
}
\end{table*}

\clearpage

\clearpage

\clearpage


\begin{thebibliography}{100}
\bibitem{note_star}
{I}n our notation ${Q}^{*}$ means "uniquely produced numerical {B}orn ${Q}$
  matrix" (with a given numerical method), consistent with a given geometry and
  ground state electronic charge density distribution function. {I}n the same
  way $\epsilon$ without indices means "uniquely produced numerical high
  frequency dielectric tensor".
\end{thebibliography}
\end{document}